\documentclass[%
 reprint,
superscriptaddress,
nofootinbib,
 amsmath,amssymb,
 aps,
 prd,
floatfix,
]{revtex4-2}

\usepackage{graphicx}
\usepackage{dcolumn}
\usepackage{bm}

\usepackage{caption,subcaption}
\usepackage{tensor}
\usepackage{titlesec}
\usepackage{xcolor}
\usepackage{ytableau}
\usepackage[colorlinks,linkcolor=blue]{hyperref}
\usepackage{ytableau}
\usepackage{orcidlink}
\usepackage{academicons}
\definecolor{orcidlogocol}{HTML}{A6CE39}
\newcommand{\orcid}[1]{\href{https://orcid.org/#1}{\textcolor[HTML]{A6CE39}{\aiOrcid}}}

\usepackage{booktabs}
\usepackage[title]{appendix}

\begin{document}

\preprint{APS/123-QED}

\title{Perturbative black-hole and horizon solutions in gravity with explicit spacetime-symmetry breaking}

\author{Samy Aoulad Lafkih \orcidlink{0009-0007-7652-8134}}
\email{samy.aoulad-lafkih@obspm.fr}
\affiliation{%
LTE, Observatoire de Paris, Université PSL, Sorbonne Universit\'e, CNRS, LNE, 61 avenue de l’Observatoire, 75014 Paris, France
}%
\author{Nils A. Nilsson \orcidlink{0000-0001-6949-3956}}%
\email{nilsson@ibs.re.kr}%
\thanks{This author contributed equally with the first author.}
\affiliation{Cosmology, Gravity and Astroparticle Physics Group, Center for Theoretical Physics of the Universe,
Institute for Basic Science, Daejeon 34126, Korea.}
 \affiliation{%
LTE, Observatoire de Paris, Université PSL, Sorbonne Universit\'e, CNRS, LNE, 61 avenue de l’Observatoire, 75014 Paris, France
}%
  \author{Marie-Christine Angonin \orcidlink{0000-0001-6000-7122}}%
 \email{marie-christine.angonin@obspm.fr}
  \author{Christophe Le Poncin-Lafitte \orcidlink{0000-0002-3811-1828}}%
 \email{christophe.leponcin-lafitte@obspm.fr}
\affiliation{%
LTE, Observatoire de Paris, Université PSL, Sorbonne Universit\'e, CNRS, LNE, 61 avenue de l’Observatoire, 75014 Paris, France
}%

\date{\today}

\begin{abstract}
In this paper, we present static and spherically symmetric vacuum solutions to the mass-dimension $d\leq 4$ action of an effective-field theory, choosing the diffeomorphism symmetry to be broken explicitly. By using the reduced-action method with a Schwarzschild seed-solution, we find static and spherically symmetric black hole solutions to the field equations to linear order in the symmetry-breaking coefficients, which are consistent solutions to the modified Einstein equations at the same order. Using several ans{\"a}tze for the symmetry-breaking coefficients we classify the allowed solutions, and we compute standard consequences and observables, including horizons, thermodynamics, photon geodesics, and perihelion precession. We find that the horizon structure of some of our solutions are similar to the Reissner-Nordstr{\"o}m case, and that several of them exhibit physical singularities at $r=2M$, as well as non-zero angular deficit at $r\to\infty$. We note in particular that introducing more than one non-zero coefficient for spacetime-symmetry breaking leads to a solution with three horizons; the aim is to obtain observables that can be confronted to black-hole observational data. 
\end{abstract}

\maketitle


\section{Introduction}\label{sec:intro}
General Relativity (GR) and the Standard Model of particle physics (SM) are successful field theories which together generate a remarkably accurate description of known physics. Conversely, it is generally expected that GR and the SM are not the ultimate descriptions of Nature, and that they will eventually be superseded by some all-encompassing theory~\cite{Weinberg:2009bg}. In this sense, GR and the SM can be viewed as low-energy effective-field theory (EFT) descriptions of such a fundamental theory, which includes a consistent quantum formulation of the gravitational interaction.\\
GR has held up exceptionally well under experimental tests for over a century.
However, many of these tests have taken place in weak gravity regime \cite{Will:2014kxa}, and 
any deviation from GR is expected to arise in strong gravity environments, where observations are more challenging. Naturally, some of the prime candidates for such tests are compact objects, such as black holes and neutron stars. 

A very complete review of observational techniques and analysis of black holes has been done by Genzel, Eisenhauer and Gillessen \cite{genzel2024experimental}. GRAVITY is an instrument within the Very Large Telescope Interferometer (VLTI) whose performances allow it to observe the close proximity of the compact object producing the strongest gravitational field in our galaxy: Sagittarius A* (supermassive black hole at the center of the Milky Way). The many teams working in collaborations on the instrument's data managed to constrain its gravitational redshift through the spectrum of the closely orbiting star S2~\cite{abuter2020detection}. More recently, they also followed its orbit around its periastron~\cite{2024arXiv240912261T,grould2017general, GRAVITY:2020gka}, thus offering new insights on its precession, an observable famously related to relativistic effects~\cite{einstein1979explanation}.   As opposed to GRAVITY that is a solitary instrument, the Einstein Horizon Telescope (EHT) is an international collaboration involving many observatories, whose goal is to cross-correlate the signals from far-away instruments in order to create an Earth-sized interferometer~\cite{Fish:2016jil}. Thanks to this method, the collaboration has been aiming for a small enough angular resolution so that it may be able to observe the regions of strong gravity near black holes. It has succeeded in doing so by observing the emission regions of the two black holes with largest apparent event horizons : Sagittarius A*~\cite{EventHorizonTelescope:2022wkp,EventHorizonTelescope:2022apq, EventHorizonTelescope:2022wok, EventHorizonTelescope:2022exc, EventHorizonTelescope:2022urf, EventHorizonTelescope:2022xqj, EventHorizonTelescope:2024hpu, EventHorizonTelescope:2024rju, EventHorizonTelescope:2023gtd} and M87~\cite{EventHorizonTelescope:2019dse, EventHorizonTelescope:2019uob, EventHorizonTelescope:2019jan, EventHorizonTelescope:2019ths, EventHorizonTelescope:2019pgp, EventHorizonTelescope:2019ggy, EventHorizonTelescope:2021bee, EventHorizonTelescope:2021srq} (located in the center of the Virgo A galaxy). But electromagnetic signals are not the only way we may probe compact objects: gravitational waves are new astrophysical messengers that should have the potential for highly accurate tests of the nature of black holes~\cite{Colleoni:2024lpj}. Thanks to current detectors (Ligo,Virgo and Kagra (LVK)) and future ones (LISA in 2035~\cite{2017arXiv170200786A}, Einstein Telescope in 2035 ~\cite{Maggiore:2019uih} and Cosmic Explorer~\cite{Reitze:2019iox} around 2037), we are about to obtain a great many numbers of gravitational waves signals (LVK collaborations has already confirmed around 90 events, see \cite{LIGOScientific:2020ibl,LIGOScientific:2021usb, KAGRA:2021vkt}), allowing us to probe compact objects through the generation of gravitational waves.

When searching for a unifying theory, it may be necessary to relax some of the canonical requirements normally demanded from a well-behaved field theory, such as local Lorentz symmetry and diffeomorphism symmetry; indeed, some proposals of quantum gravity predict that local Lorentz symmetry are no longer exact at the energy scales relevant for a quantum description of gravity~\cite{Kostelecky:1988zi,Kostelecky:1991ak,Colladay:1996iz,Colladay:1998fq,Gambini:1998it,Carroll:2001ws,Addazi:2021xuf,Mariz:2022oib,Horava:2009uw}. Such {\it spacetime-symmetry breaking} has attracted significant attention in the literature in the past decades, computing observables that could be confronted to observations, and the current interest is high~\cite{Bailey:2022wuv,Petrov:2020wgy,Safronova:2017xyt,Will:2014kxa,Tasson:2014dfa,Liberati:2013xla,Kostelecky:2008ts}.
There is also a large interest in the study of specific models and non-EFT approaches to spacetime-symmetry breaking, see for example~\cite{Amelino-Camelia:2008aez,Mattingly:2005re,Wang:2017brl,Skordis:2020eui}.

One way to introduce spacetime-symmetry breaking is to place a fixed background tensor at the level of the Lagrangian; this approach is called {\it explicit breaking} and has been described in detail in~\cite{Bluhm:2014oua}. Different complications, including Bianchi-identity compatibility, arise in such models which nevertheless have rich and interesting phenomenological consequences. These have been considered in for example \cite{Kostelecky:2003fs,Kostelecky:2020hbb,ONeal-Ault:2020ebv,Nilsson:2022mzq,Khodadi:2023ezj,Reyes:2022mvm}. One interesting avenue for constraining such explicit-breaking models is to construct solutions for compact objects such as black holes, which is exactly the topic of the present work: in this paper, we will construct static and spherically symmetric vacuum solutions where particle diffeomorphisms are explicitly broken by a two-tensor, which is currently not present in the literature. In the context of the EFT adopted in this paper, solutions have also been found in explicit breaking scenarios in terms of a 4-tensor coefficient in~\cite{Bonder:2020fpn}. Similar approaches have previously been considered in the spontaneous-breaking case, for example~\cite{Xu:2022frb,Casana:2017jkc,Lessa:2019bgi,AraujoFilho:2024ykw}, and several aspects and classical tests of such solutions were presented in for example~\cite{Uniyal:2022xnq,Liang:2022hxd,Khodadi:2022mzt,Tuleganova:2023izp,Xu:2023xqh,Bonder:2021gjo,Bertolami:2005bh}, and references therein. Interestingly, the authors in \cite{khodadi2022probing} recently placed constraints using the images of Sagittarius A* and M87* and studying the impact of non-zero EFT coefficients on the size of the shadow.

The purpose of this paper is to obtain and analyse some simple spherically symmetric vacuum solutions for a 2-tensor EFT coefficient coupled to the Ricci tensor as a correction to GR~\cite{Colladay:1996iz,Colladay:1998fq} in the context of explicit diffeomorphism violation. Rather than considering the coefficient in full generality, we adopt the ``one-at-a-time''\footnote{When similar techniques are used for experimental constraints, they are known as ``maximun-reach'' methods \cite{Kostelecky:2008ts}.} method where we only let one tensor component at a time take on a non-zero value, after which we combine our solutions for a more general result. Such techniques, introduced for simplicity, have previously been used in the literature, for example in \cite{Bailey:2024zgr}. Since any explicit diffeomorphism breaking has been shown to be small \cite{Nilsson:2022mzq,Kostelecky:2021tdf}, we introduce a linear perturbation scheme around the GR solution, and we obtain Schwarzschild at zeroth order along with symmetry-breaking corrections. This approach is similar to the one used in \cite{Bonder:2020fpn}, where the authors adopt a perturbative scheme and find corrections to the Schwarzschild metric using a 4-index tensor; however, since we do not consider traceless coefficients in this paper, our results are not subsets of \cite{Bonder:2020fpn}.

The paper is organised as follows: in Section~\ref{sec:EFT} we introduce the EFT framework used to generate symmetry-breaking corrections to GR, highlighting the no-go constraints which generally must be avoided in the case of explicit breaking; in Section~\ref{sec:vacsols} we derive several new perturbative vacuum solutions to the modified Einstein equations; in Section~\ref{sec:implications} we study the near-horizon behaviour and thermodynamics of the new solutions; in Sections~\ref{sec:photonradialgeodesics}, \ref{sec:consquantsorbits}, \ref{sec:precession} we study photon radial geodesics, conserved quantities and orbits, and periaston precession implied by our solutions; in Section~\ref{sec:lightring}, we discuss the dynamics of the light ring; in Section~\ref{sec:disc} we discuss and conclude. Throughout this paper, we adopt $c=\hbar=1$, except in Section~\ref{sec:photonradialgeodesics} where we restore $c$ temporarily; moreover, we adopt the $(-+++)$ metric signature.

\section{Effective-field theory}\label{sec:EFT}

Diffeomorphism-breaking EFT corrections to the Einstein-Hilbert action are constructed by contracting EFT coefficients with curvature operators and their covariant derivatives. The mass-dimension $d\leq4$ EFT action we consider can be written as~\cite{Colladay:1996iz,Colladay:1998fq,Kostelecky:2003fs}
\begin{equation}\label{eq:lagr}
    S[g] = \frac{1}{2\kappa}\int d^4x \sqrt{-g}\left(R+(k_R)^{\mu\nu\alpha\beta}R_{\mu\nu\alpha\beta}\right)+S',
\end{equation}
where $\kappa=8\pi G$, $R$ and $R_{\mu\nu\alpha\beta}$ is the curvature scalar and the Riemann tensor, respectively. $(k_R)^{\mu\nu\alpha\beta}$ is an EFT coefficient which in this case is considered a fixed background tensor under particle transformations\footnote{Such transformations are outlined in detail elsewhere, see for example \cite{Kostelecky:2020hbb}.}; $(k_R)^{\mu\nu\alpha\beta}$ inherits the same index symmetries as the Riemann tensor. Lastly, $S'$ contains dynamical terms and matter terms, which we set to zero as we are considering explicit breaking and vacuum. For a particle diffeomorphism 
 along $\xi^\mu$, the background tensor does not transform, i.e. the term $\mathcal{L}_\xi k_R$ does not appear in the transform, which therefore explicitly breaks the particle diffeomorphism invariance of the action $S[g]$\footnote{Local Lorentz invariance is still preserved; see Table II in \cite{Kostelecky:2020hbb} for a general classification scheme.}. The EFT correction can be expanded using a Ricci decomposition of the Riemann tensor to read
\begin{equation}\label{eq:riccidecomp}
    (k_R)^{\mu\nu\alpha\beta}R_{\mu\nu\alpha\beta} = s^{\mu\nu}R_{\mu\nu}+t_{\mu\nu\alpha\beta}C^{\mu\nu\alpha\beta},
\end{equation}
where $R_{\mu\nu}$ is the Ricci tensor and $C^{\mu\nu\alpha\beta}$ is the Weyl tensor, and where the EFT coefficients $s$ and $t$ inherit the index symmetries of their parent curvature tensors. Note that it is possible to further decompose the first term into a saturated trace and a trace-free two-tensor piece. The case of $t_{\mu\nu\alpha\beta}$ has been considered elsewhere~\cite{Bonder:2020fpn}: we set it to zero in this paper, and focus on the effects of the symmetric two-tensor $s^{\mu\nu}$ (we make no traceless assumption here); as such, the Lagrange density reads 
\begin{equation}\label{eq:simpleL}
S_1[g]=\frac{1}{2\kappa}\int d^4x\sqrt{-g}(R+s^{\mu\nu}R_{\mu\nu}).
\end{equation}

By varying the resulting action with respect to the metric and considering the fully contravariant tensor $s^{\mu\nu}$ as the independent field (i.e. $s_{\mu\nu} := g_{\mu\alpha}g_{\nu\beta}s^{\alpha\beta}$), we obtain the modified Einstein equations as
\begin{equation}\label{eq:Eeqs}
    \begin{split}
         G^{\alpha\beta}-&\tfrac{1}{2}g^{\alpha\beta}R^{\mu\nu}s_{\mu\nu}+\tfrac{1}{2}g^{\alpha\beta}\nabla_\mu\nabla_\nu s^{\mu\nu} \\
         &-\nabla_\mu\nabla^{(\alpha}s^{\beta)\mu}+\tfrac{1}{2}\nabla_\mu\nabla^\mu s^{\alpha\beta} = 0,
    \end{split}
\end{equation}
and from the associated traced Bianchi identities we obtain
\begin{equation}\label{eq:Bianchi}
    \tfrac{1}{2}R^{\mu\nu}\nabla^\beta s_{\mu\nu} + \nabla_\nu\left(s^{\mu\nu}R^\beta_{~\mu}\right) = 0,
\end{equation}
which requires a judicious choice of $s^{\mu\nu}$, since we lack any other constraint equation. For example, it was found in \cite{ONeal-Ault:2020ebv} that a homogeneous coefficient ($\partial_i s^{\mu\nu}=0$) with a modified continuity equation of state for radiation and cosmological constant allowed for a consistent solution using the flat FLRW metric.

\subsection{A trace subset}
We also investigate a scalar-type correction to GR of the form
\begin{equation}\label{eq:lagtrace}
    S_{\rm trace}[g]=\frac{1}{2\kappa}\int d^4x \sqrt{-g}R\left(1+s^\mu_{~\mu}\right),
\end{equation}
which was recently considered in an extension to the minimal EFT in \cite{Bailey:2024zgr}, and was shown to contain an extra degree of freedom contained in the trace.
We note that this action is not equivalent to introducing the trace of $s^{\mu\nu}$ as a scalar quantity due to the extra metric dependence. As in the previous case, we will consider $s^{\mu\nu}$, i.e. the fully contravariant indices as being ``independent''; the Einstein equations in this case read
\begin{equation}\label{eq:Eeqstrace}
    \begin{aligned}
        G^{\alpha\beta}&\left(1+s^\mu_{~\mu}\right)+R s^{\alpha\beta} +\left(g^{\alpha\beta}\nabla_\lambda\nabla^\lambda - \nabla^\beta\nabla^\alpha\right)s^\mu_{~\mu} = 0,
    \end{aligned}
\end{equation}
where one more term is now present compared to the scalar trace case. We find the Bianchi identities as
\begin{equation}
    \tfrac{1}{2}R\nabla^\beta s^\mu_{~\mu} + \nabla_\alpha\left(s^{\alpha\beta}R\right) = 0,
\end{equation}
which closely resemble those in Eq.~(\ref{eq:Bianchi}). We note that this trace subset was considered in \cite{Bailey:2024zgr} where it was found that it leads to an unsuppressed (a factor of $3$ compared to GR) gravitational scalar mode; however, in \cite{Bailey:2024zgr}, the covariant components of $s_{\mu\nu}$ were chosen to be fixed under variation and therefore constitutes a different model.

\section{New vacuum solutions}\label{sec:vacsols}
In this section, we derive simple spherically symmetric and static vacuum solutions satisfying the modified Einstein equations (\ref{eq:Eeqs}) and (\ref{eq:Eeqstrace}).
We begin by choosing the metric ansatz
\begin{equation}\label{eq:ansatz}
    (ds^2)_{\rm ansatz} = -N^2(r)dt^2 + \tfrac{1}{f(r)}dr^2 + r^2 d\Omega_2^2,
\end{equation}
where $N(r)$ and $f(r)$ are unknown functions of the radial coordinate $r$ and $d\Omega_2$ is the solid-angle element of a 2-sphere. To find $N(r)$ and $f(r)$ which solves the Einstein equations from the previous section, we use Palais' symmetric criticality~\cite{Palais:1979rca}, or the ``reduced-action method'': although widely used, it is worth noting that in our case, the diffeomorphism symmetry group of GR has been explicitly broken, and our Lagrangians are only invariant under the observer subgroup. Thankfully, symmetric criticality is applicable also to such non-covariant theories, as shown in \cite{Fels:2001rv}, and we are free to proceed with some caution. Given the details of the theory, the existence of non-trivial boundary terms may appear during the reduced variational procedure, and it is necessary to perform careful consistency checks of the reduced solutions.

Given the symmetries of our metric ansatz $(ds^2)_{\rm ansatz}$ and the structure of the resulting Einstein equations, we choose the coefficient tensor $s^{\mu\nu}$ to be diagonal, i.e. in some basis, it reads
\begin{equation}
    s^{\mu\nu} \rightarrow
        \begin{pmatrix}
            s^{00} & 0 & 0 & 0 \\
            0 & s^{11} & 0 & 0 \\
            0 & 0 & s^{22} & 0 \\
            0 & 0 & 0 & s^{33}
        \end{pmatrix},
\end{equation}
and from now one, we consider the components of $s^{\mu\nu}$ to be constants in the chosen coordinate system, i.e. $\partial_\mu s^{\alpha\beta}=0$. In this paper, we will only consider $\{s^{00}, s^{11}\}\neq0$, as the Einstein equations will demand (due to spherical symmetry) that the coefficients $s^{22}$ and $s^{33}$ vanish at the background level.\footnote{If we were to consider general perturbations around the metric (\ref{eq:ansatz}), we would need to account for these terms.}

\subsection*{Case 1: $s^{00}\neq 0$}\label{sec:case1}
Starting with the case when only $s^{00}\neq0$, we plug in the metric ansatz (\ref{eq:ansatz}) into the Lagrangian in Eq.~(\ref{eq:simpleL}) and find the following reduced action (after angular integration)
\begin{equation}
    \begin{aligned}
        S_{1} =& \frac{1}{2\kappa}\int dr \Bigg[-\frac{2}{\sqrt{f}}\Big[2N(f+rf'-1)+r(N'(4f+rf') \\&+2fN'') \Big] + s^{00}\frac{N^2}{\sqrt{f}}r\left[N'(4f+rf')+2rfN''\right] \Bigg],
    \end{aligned}
\end{equation}
where primes denote differentiation with respect to the radial coordinate $r$. We find the Euler-Lagrange equations for $N(r)$ and $f(r)$ as
\begin{equation}\label{eq:ELcase1}
    \begin{aligned}
            0{}=&{}1 +rf' \left( -1 + s^{00}NrN' \right) + f ( -1 + s^{00} r^2 N'^2 \\
            &+ 2s^{00} Nr \left( 2N' + rN'' \right) ), \\
    0{}=&{}-2frN' + N \left( 1 + f \left(-1 + s^{00} r^2 N'^2 \right) \right), \\
    \end{aligned}
\end{equation}
which we now solve using an order-by-order prescription (see Appendix~\ref{app:simpleexact} for an outline of simple exact solutions of these non-linear differential equations). First, we split $N(r)$ and $f(r)$ into a zeroth-order term (this will turn out to be the Schwarzschild solution) plus a correction linear in the EFT coefficient $s^{00}$. In this way, the zeroth-order term acts as a seed solution for first order; by adding the zeroth and perturbation-order solutions together, we arrive at expressions for $N(r)$ and $f(r)$ which satisfy both the Euler-Lagrange equations (\ref{eq:ELcase1}) and the covariant Einstein equations (\ref{eq:Eeqs}) at linear order (after consistency checks are made).

Expanding the dynamical variables as $N(r) \to N_{0}(r)+N_{1}(r)$ and $f(r) \to f_{0}(r)+f_{1}(r)$, we write the zeroth-order equations in Eq.~(\ref{eq:ELcase1}) as
\begin{equation} \label{eq:eqdiff1}
    \begin{aligned}
   0{}&= 1 - rf'_0 - f_0  \\
    0{}&=-2f_0 rN'_0 + N_0 \left( 1 - f_0 \right),
    \end{aligned}
\end{equation}
for which we immediately find the Schwarzschild solution
\begin{equation}\label{eq:schwarz}
     f_0 = N_0^2 = 1 + \frac{k}{r},
\end{equation}
where $k$ is a negative mass parameter, with $k\to -2M$ corresponding to Schwarzschild, but other choices are possible, leading to horizonless objects and other exotic solutions. For the rest of this paper, we keep $k$ fully generic (albeit negative) and when comparing to the Schwarzschild solution, we set $k\to-2M$.

Plugging the above solutions into Eq.~(\ref{eq:ELcase1}) gives us the equations for the first-order corrections as
\begin{equation}
    \begin{aligned}
      0={}& 1 - f_1 - rf'_1 +rf'_0 ( -1 + s^{00} N_{0} rN'_{0} ) \\&+ f_0 ( -1 + s^{00} r^2 (N'_{0})^{2} + 2s^{00} N_0 r \left( 2N'_0 + rN''_0 \right) ),  \\
    0= {}& 2r f_0 N'_1 + N_1 (f_0 - 1) + 2r (f_1 N'_0 + f_0 N'_0) \\&- N_0 ( 1 + f_0 (-1 + s^{00} r^2 (N'_{0})^{2} ) ) + N_0 f_1.
    \end{aligned}
\end{equation}
The first equation can be solved for $f_1(r)$ with no further input, after which we plug the solution into the second equation. We find that the first-order corrections to the metric functions are
\begin{equation}
    \begin{aligned} \label{eq:solpart1}
        f_1 ={}& - s^{00} \frac{k^2}{4 r^2} + \frac{c_1}{r} \\
        N_1 ={}&  - s^{00} \frac{k^2}{8 r^2} \left( 1 + \frac{k}{r} \right)^{-\frac{1}{2} } + c_2 \left( 1 + \frac{k}{r} \right)^{\frac{1}{2}},
    \end{aligned}
\end{equation}
where $c_1$ and $c_2$ are integration constants. We may put $c_1 \to 0$ by noticing that it is simply a scaling of the zeroth-order solution, and $c_2$ needs to vanish by demanding a smooth GR limit, i.e. that $N_1(r) \to 0$ as $s^{00} \to 0$; this choice also avoids a vDVZ-type discontinuity which has recently been shown to appear in explicit-breaking scenarios \cite{Bailey:2024zgr}. Other choices of boundary values are possible; for example, if we choose $c_2 \propto s^{00}$, we obtain a different solution which respects the GR limit.

Finally, we write down our full solution for the static and spherically symmetric metric as
\begin{equation}\label{eq:scase1}
\boxed{%
    \begin{aligned}
        N(r) ={}&\left(1+\frac{k}{r}\right)^{\tfrac{1}{2}} - s^{00} \frac{k^2}{8 r^2} \left( 1 + \frac{k}{r} \right)^{-\tfrac{1}{2}} \\
        f(r) ={}&1+\frac{k}{r} - s^{00} \frac{k^2}{4 r^2},
    \end{aligned}
    }
\end{equation}
which satisfies the Einstein equations (\ref{eq:Eeqs}), traced Bianchi identities (\ref{eq:Bianchi}), as well as $N^2(r)/f(r) = 1 + \mathcal{O}(s^{00})^2$ up to first order in the EFT coefficient. Moreover, $f(r) \to 0 + \mathcal{O}(s^{00})^2$ at the horizon, whereas $N(r) \to -s^{00}/4 + \mathcal{O}(s^{00})^2$.

We may attempt to verify this solution using an exact method. Since $N^2=f$ holds perturbatively to first order in $s^{00}$, we backtrack to Eq.~(\ref{eq:ELcase1}) and plug in the ansatz $N^2=f$, after which we obtain
\begin{equation}
    f+rf'-\frac{s^{00}}{4}r^2f'^2-1 = 0,
\end{equation}
without using the order-by-order prescription. This equation has an exact solution in terms of Lambert functions $W(x)$\footnote{Solution to $x=W(x)e^{W(x)}$ and with derivative $W'(x)=\frac{W(x)}{x(1+W(x))}$\cite{Siewert:1973:EAS}.} as
\begin{equation}\label{eq:fsolLambert}
    f(r) = 1 + \frac{2W(\tfrac{C}{r})+W(\tfrac{C}{r})^2}{s^{00}},
\end{equation}
where $C$ is an arbitrary constant. Using properties of $W(x)$, it can easily be verified that $f(r)\to 1$ as $r\to\infty$. $f(r)$ appears to be singular in $s^{00}$, and there is thus no GR limit to the solution; moreover, since the constant $C$ is undetermined, there is no guarantee that $C/r$ lies within the convergence radius of $W(x)$ ($x \in (-1/e, +\infty)$). In order to cure the singularity in $f(r)$, we need to determine $C$ such that $(s^{00})^{-1} \left(2W(C/r) + W(C/r)^2 \right) \to k/r$ as $s^{00}\to 0$. Using the Taylor series of the Lambert function around $C=0$ and assuming that $C$ is linear in $s^{00}$, we find 
\begin{equation}\label{eq:fsolLambert1order}
    f(r) = 1+\frac{2\tfrac{C}{r}+\mathcal{O}((s^{00})^2)}{s^{00}},
\end{equation}
and in order for us to obtain the Schwarzschild solution (\ref{eq:schwarz}), it is necessary to fix $C$ as $C=s^{00} k/2$. Using this expression for $C$, we go to second order in the Taylor expansion (\ref{eq:fsolLambert1order}) and find
\begin{equation}
    f(r) = 1+\frac{k}{r}-s^{00}\frac{k^2}{4r^2} + \mathcal{O}((s^{00})^2),
\end{equation}
which exactly coincides with our perturbative solution in Eq.~(\ref{eq:scase1}) to first order; the solution (\ref{eq:scase1}) is therefore a linear subset of the full solution. We note however, that the exact solution (\ref{eq:fsolLambert}) with $N^2=f$ does not respect the (non-linearised) Einstein equations, and although $N^2=f$ holds to first order in the coefficients, we expect a deviation ($N^2=f+\delta f$) at some higher order, where the simple properties of $N^2=f$ spacetimes are lost.

\subsubsection{Trace subset solution}
 Plugging in the same metric ansatz as in the previous sections into the Lagrangian (\ref{eq:lagtrace}) and setting the matter sector to zero, we arrive at the following reduced action
\begin{equation}
    \begin{aligned}
    S_{\rm trace} = \frac{1}{2\kappa}\int dr &\frac{1}{\sqrt{f}}\Big[2\left(s^{00}N^2-1\right)(2N(f+rf'-1) \\&+r(rf'N'+2f(2N'+rN'')))\Big],
    \end{aligned}
\end{equation}
from which we find the Euler-Lagrange equations for $N(r)$ and $f(r)$ as
\begin{equation}
\label{eq:Eul-Lag1}
\begin{aligned}
0 =& -4\left(f+rf'-1\right)+s^{00}(12N^2(f+rf'-1) +8r^2f(N')^2 \\&+8rN\left(rN'f'+2f(2N'r^2 N''\right)), \\
0 =& 2N(f-1)+4rfN'+2Ns^{00}(N^2(f-1) +6rfNN'\\&+2r^2f(N')^2).
\end{aligned}
\end{equation}

Introducing as before $N = N_0 +N_1$ and $f=f_0+f_1$, where the zeroth order is the Schwarzschild solution, we arrive at the following static and spherically symmetric spacetime for the trace subset (see subsection \ref{subsec:appcase1tracesol} in the Appendix for a description of the simple exact solutions of the non-linear differential equations)
\begin{equation}
\label{eq:tracesol}
\boxed{%
\begin{aligned}
    N(r) &= \left(1+\frac{k}{r}\right)^{\tfrac{1}{2}} + s^{00}\frac{3k^2+4kr}{4r^2}\left(1+\frac{k}{r}\right)^{-\tfrac{1}{2}} \\
    f(r) &= 1+\frac{k}{r} -s^{00}\frac{k^2}{2r^2}.
\end{aligned}}
\end{equation}

\subsubsection*{Case 1: Numerical verification}\label{sec:num}
In order to verify the solutions obtained in the previous Sections, we construct here a set of numerical solutions of the functions $N(r)$ and $f(r)$. The Euler-Lagrange equations in (\ref{eq:Eul-Lag1}) are highly non-linear, but fortunately we can still extract an expression for the derivatives of $f$ and $N$ that only depends on local parameters $r$, $f(r)$, $N(r)$ and $s^{00}$ as 
\begin{figure}
    \centering
    \begin{subfigure}[T]{0.49\textwidth}
        \includegraphics[width=\textwidth]{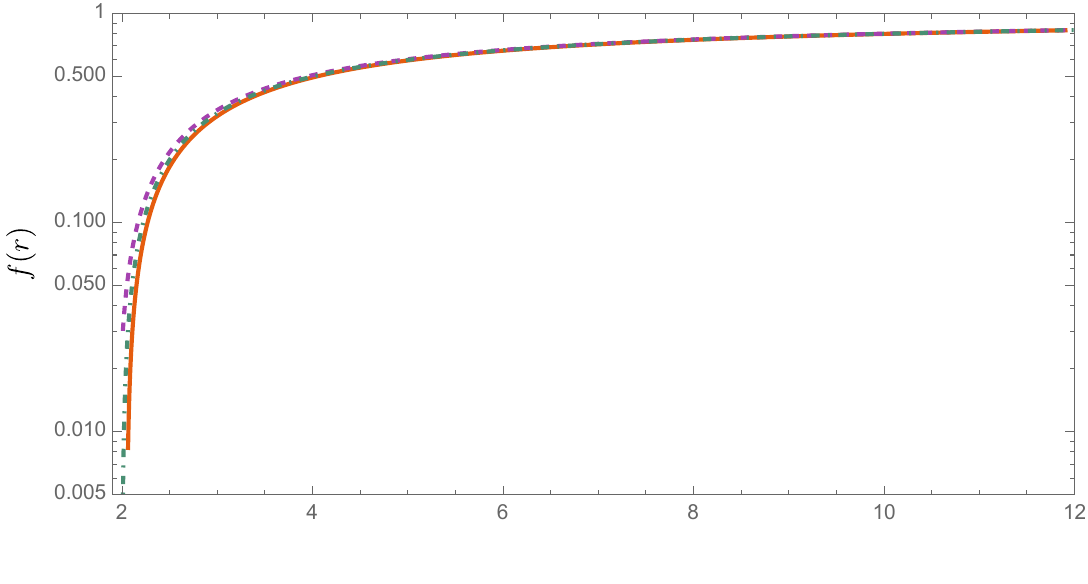}
        \caption{Numerical solution for $f$.}
        \label{fig:numintegf}
    \end{subfigure}
    \hfill
        \begin{subfigure}[T]{0.49\textwidth}
        \includegraphics[width=\textwidth]{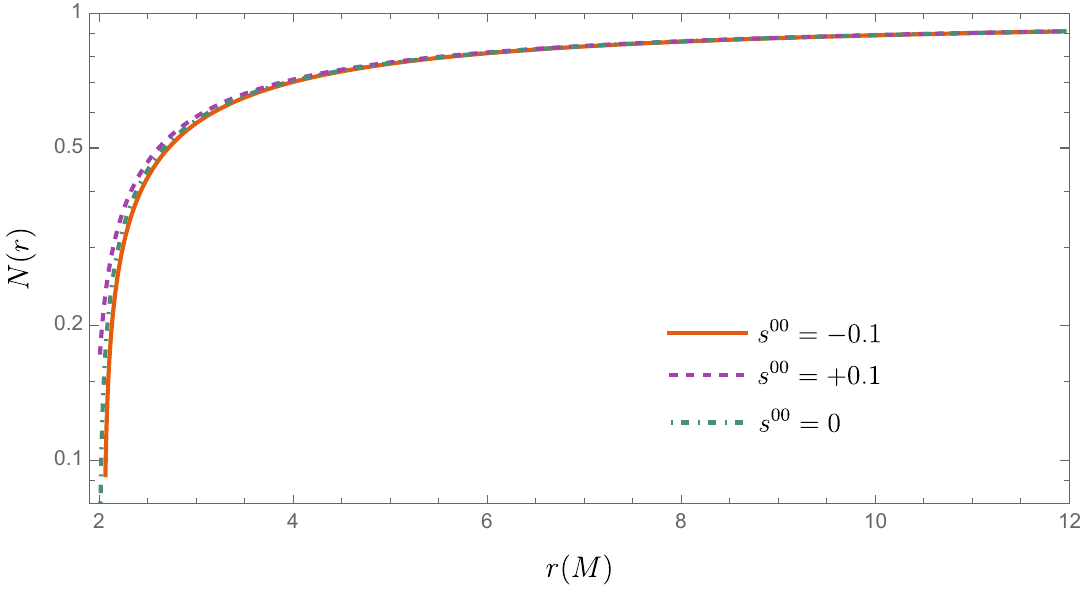}
        \caption{Numerical solution for $N$.}
        \label{fig:numintegN}
    \end{subfigure}
    \caption{The metric functions $f$ (top) and $N$ (bottom) for Case 1 numerically integrated (solid lines, Eq.~(\ref{eq:ELcase1})), as well as the GR limit (dashed lines).}
    \label{fig:numinteg}
\end{figure}
and we note that there exists a solution $(f'(r), N'(r))$ to the system (\ref{eq:ELcase1}) on the condition that f(r) 
\begin{equation}
\left( f(r) + s^{00} N(r)^2 \left(f(r) - 1 \right) \right) \geqslant 0.
\end{equation}
This allows us to extrapolate the next point at $r + \Delta r$ knowing only the value of the functions at $r$ using the Euler scheme
\begin{figure}
    \centering
    \begin{subfigure}[T]{0.49\textwidth}
        \includegraphics[width=\textwidth]{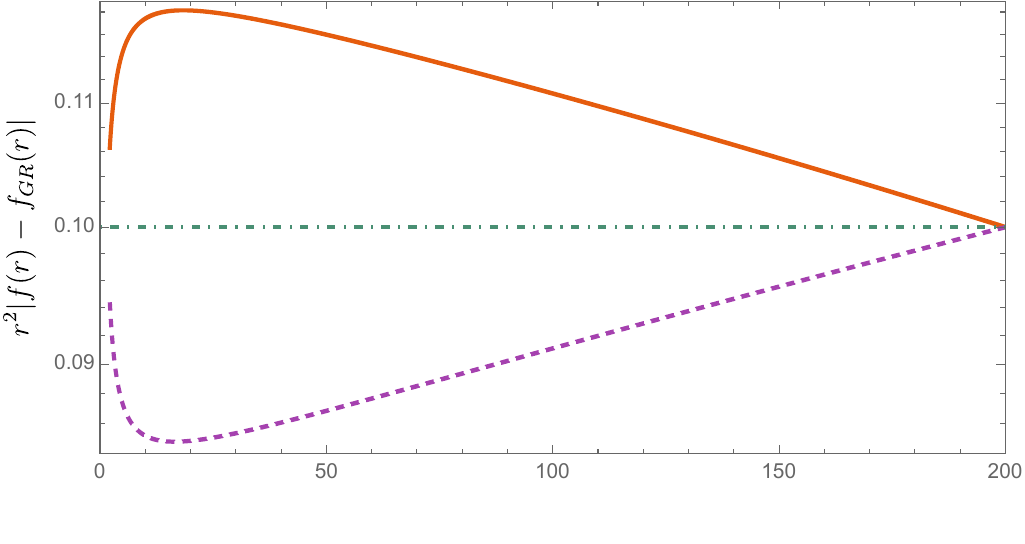}
        \caption{Comparison of $f$ and the Schwarzschild case.}
        \label{fig:numintegf}
    \end{subfigure}
    \hfill
        \begin{subfigure}[T]{0.49\textwidth}
        \includegraphics[width=\textwidth]{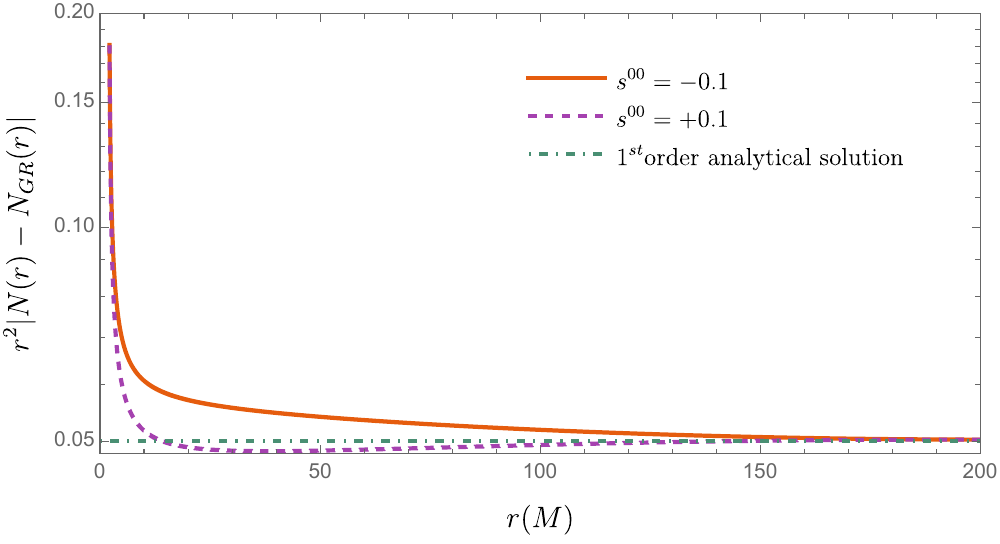}
        \caption{Comparison between $N$ and the Schwarzschild case.}
        \label{fig:numintegN}
    \end{subfigure}
        \caption{Comparison between integrated metric (solid lines, Eq.~(\ref{eq:ELcase1})) functions $f$ (top) and $N$ (bottom), and the analytical 1st order solution (dashed lines, Eq.~(\ref{eq:scase1})).}
        \label{fig:numinteg2}
\end{figure}

Figure~\ref{fig:numinteg} shows the difference between this numerical integration and the Schwarzschild solution, for an exaggerated value of $s^{00} = \pm 0.1$. We remark that when $r$ reaches a few hundred Schwarzschild radii, the metric functions become very close to the GR limit. Even with an exaggeratedly high value of $s^{00}$ like $0.1$, the different cases diverge only slightly, even close to the event horizon.

We also compare the analytical first order solution with the numerically integrated metric, which is shown in Figure~\ref{fig:numinteg2}. We see that there is good agreement between the analytical solution and the integration up to the Schwarzschild horizon at $r=2M$.

\subsection*{Case 2: $s^{11} \neq 0$}\label{sec:case2}
Repeating the same exercise as in the previous section, but now letting $s^{11} \neq 0$, we obtain the following solution to first order in $s^{11}$ (please see subsection \ref{subsec:appcase2sol} in the Appendix for a description of the simple exact solutions of the non-linear differential equations)
\begin{equation}\label{eq:scase2}
\boxed{%
    \begin{aligned}
        N(r) =& \left(1+\frac{k}{r}\right)^{\tfrac{1}{2}} + s^{11}\frac{k}{8r}\left(4+5\frac{k}{r}\right)\left(1+\frac{k}{r}\right)^{-\tfrac{3}{2}} \\
        f(r) =& 1+\frac{k}{r} + s^{11}\left[\frac{3k^2}{4r^2}\left(1+\frac{k}{r}\right)^{-1}-1\right]
    \end{aligned}
    }
\end{equation}
In this case we observe that the metric does not tend towards Minkowski at large $r$; indeed, according to (\ref{eq:scase2}), when $r \to +\infty$ we find that $N(r) \to 1$ and $f(r) \to 1 - s^{11}$.\\
If we rescale the radial coordinate $r$ as $\Tilde{r} = \frac{r}{\sqrt{1-s^{11}}}$, we recover $\Tilde{f}(r) \to 1$ at spatial infinity. In order to obtain the Minkowski metric at infinity, we also need to rescale the angular coordinates by $\Tilde{\theta} =\theta \sqrt{1-s^{11}}$ and $\Tilde{\varphi} =\varphi \sqrt{1-s^{11}}$. However, doing so implies that the angular coordinates are no longer $2 \pi$-periodic, but $(2\pi\sqrt{1-s^{11}})$-periodic. In this case, we have a kink solution which hints at some type of topological defect. This was also noticed in the context of bumblebee black-hole solutions in \cite{Li:2008tma}.

\subsection*{Case 3: $s^{00} \neq 0$, $s^{11} \neq 0$}\label{sec:case3}
We also investigate the case when both $s^{00}$ and $s^{11}$ are non-zero. Due to the linearity of the problem, it suffices to sum the solutions found in the previous two subsections. 


\begin{equation}\label{eq:scase3}
    \begin{aligned}
        N(r) =& N_0 \left( r \right) + \sum_{i=1,2}[N_1(r)]_{\rm{Case}~i} \\
        f(r) =& f_0 \left( r \right) + \sum_{i=1,2}[f_1(r)]_{\rm{Case}~i}
    \end{aligned}
\end{equation}

This combined solution also satisfies the modified Einstein equations, and although finding them from the one-at-a-time cases above is simple, the presence of two non-zero coefficients leads to complications in the following sections.

\section{Horizons and thermodynamics}\label{sec:implications}
In this Section, we study the properties of the solutions found in the previous sections. A reference guide to the solutions can be seen in Table~\ref{tab:solguide}.

\renewcommand{\arraystretch}{1.2}
\setlength{\tabcolsep}{4pt}
\begin{center}
\begin{table}[h!]
\centering
    \begin{tabular}{ c c c }
    Case & $s_{\mu\nu}R^{\mu\nu}$ & $s^{\alpha}_{~\alpha}$ \\
      \midrule
      \textbf{Case 1} & Eq.~(\ref{eq:scase1}) & Eq.~(\ref{eq:tracesol}) \\
      \textbf{Case 2} & Eq.~(\ref{eq:scase2}) &  --\\
      \textbf{Case 3} & Eq.~(\ref{eq:scase3}) &  --\\\midrule
     \end{tabular}
     \caption{Guide to the four solutions considered in this paper.} 
     \label{tab:solguide}
     \end{table}
\end{center}

\subsection{Case 1}\label{sec:propcase1}

\subsubsection{Horizons}\label{sec:case1hor}

Starting from Eq.~(\ref{eq:scase1}) we find the horizon radius $r=r_\star$ by solving $f(r_\star)=0$\footnote{For this case, we have $N^2=f$ to linear order in $s^{00}$, and so we can equivalently solve $N^2(r_\star)=0$}., which reads
\begin{equation}
    r_\star^2 + kr_\star - s^{00}\frac{k^2}{4} = 0,
\end{equation}
which has two real roots since $s^{00}\ll1$ and can be written as
\begin{equation}\label{eq:horcase1}
    r_{\star, 1} = \tfrac{-k}{2}\left(1\pm \sqrt{1+s^{00}}\right),
\end{equation}
or, using that $s^{00}\ll1$
\begin{equation}
\label{eq:case1hor}
r_{\star, 1} = \frac{-k}{2}\left(1\pm \left(1+ \frac{s^{00}}{2} \right)\right) + \mathcal{O}((s^{00})^2),
\end{equation}
where the subscript $``1"$ refers to Case 1.
The positive horizon branch coincides with that of Schwarzschild for $s^{00}\to0$, but we notice that whilst the outer horizon is always present and close to the Schwarzschild radius ($r_\star=-k$), the inner lies very close to the singularity and becomes negative for $s^{00}>0$. This horizon does not coincide with that of Schwarzschild and is another example of discontinuities which have been observed in other systems with explicit diffeomorphism breaking. It may also be a spurious horizon, i.e an artifact of the perturbative nature of the solution; moreover, since this metric solution is homogeneous (i.e. $N^2 = f + \mathcal{O}\left(s^{00}\right)^2$), Eq.~(\ref{eq:case1hor}) is also a Killing horizon associated with stationarity property of the spacetime.
For the trace model in Eq.~(\ref{eq:tracesol}), we solve $1+k/r_{\star} - s^{00}k^2/2r_\star^2 = 0$ and find horizons which are the same as the above case, except that $s^{00}$ is scaled by a factor of two; however, these are not Killing horizons, as is shown in Appendix~\ref{app:killinghorizons}.

\subsubsection{Thermodynamics}\label{sec:case1thermo}
\begin{figure}
    \centering
    \begin{subfigure}[T]{0.49\textwidth}
        \includegraphics[width=\textwidth]{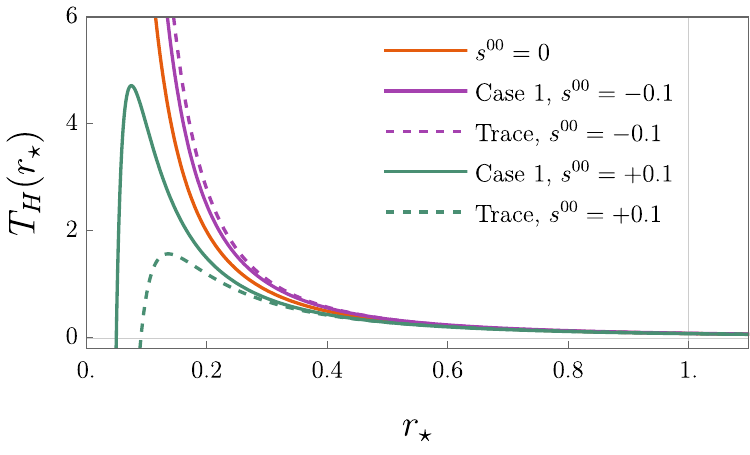}
        \caption{The Hawking temperature as a function of the outer horizon radius $r_\star$ for Case 1 and the trace model for two different values of $s^{00}$ compared with the GR limit (orange).}
        \label{fig:TH1andtrace}
    \end{subfigure}
    \hfill
        \begin{subfigure}[T]{0.49\textwidth}
        \includegraphics[width=\textwidth]{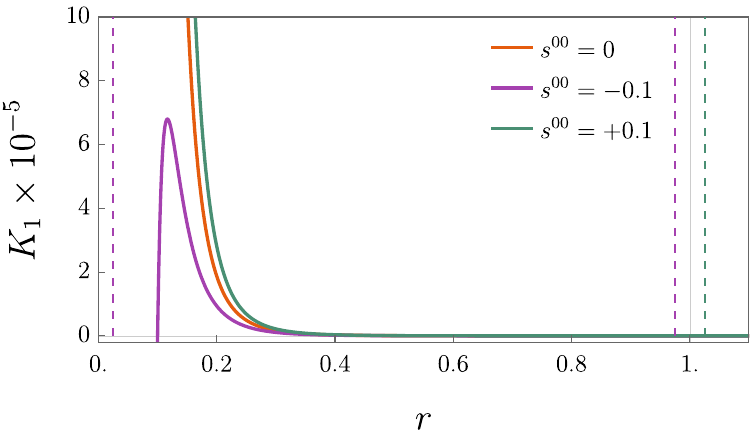}
        \caption{The Kretschmann scalar for Case 1, where the $s^{00}=0$ case is that of GR and Schwarzschild. Dashed lines indicate the horizons for the corresponding solution, and the solid grey line represents the Schwarzschild horizon.}
        \label{fig:kl1}
    \end{subfigure}
    \caption{Hawking temperature (left) and Kretchmann scalar (right) for Case 1.}
\end{figure}

In order to establish the Hawking temperature $T_H$, we may use either the quantum tunneling method or the Euclidean method\footnote{We leave computing the free energy and counterterm for future work.}, as shown in~\cite{Gomes:2018oyd}; both methods lead to
\begin{equation}\label{eq:hawkingdef}
    T_H = \tfrac{1}{4\pi}\sqrt{(N^2(r))'f'(r)}\big|_{r\to r_\star}.
\end{equation}
We compute the Hawking temperature for the solution (\ref{eq:scase1}) to Case 1 and find that
\begin{equation}
    \begin{aligned}
        T_{H,1} &= \frac{|k|}{4\pi r_\star^2}-\frac{k|k|}{8\pi r_\star^3}s^{00} + \mathcal{O}((s^{00})^2) \\
        T_{H,{\rm trace}} &= \frac{|k|}{4\pi r_\star^2}+\frac{|k|}{4\pi}\left(\frac{1}{r_\star^2}+\frac{k}{r_\star^3}\right)s^{00} + \mathcal{O}((s^{00})^2).
    \end{aligned}
\end{equation}
The above expressions indicate that the behaviour of the temperature depends on the sign of $s^{00}$, and we notice that for $s^{00}>0$, there will be a turning point at some critical radius where $T_H$ reaches a maximum value and then drops; this can be seen in Figure~\ref{fig:TH1andtrace}. We note here that the negative values of the Hawking temperature, which is canonically defined on the outermost horizon, occurs for ``small'' values of horizon radius $r_\star$, achievable only by very small masses and/or unphysical values of $s^{00}$. Nevertheless, negative Hawking temperatures have been considered in e.g.~\cite{Park:2006fp} in the case of higher-curvature black holes and other exotic solutions. Moreover, it was found in~\cite{Nozari:2007px} that the Hawking temperature can take on negative values when Lorentz symmetry is broken, or when considering non-commutative effects in the Hayward solution~\cite{Mehdipour:2016vxh}. This suggests that extensive thermodynamics may not suffice in the context of broken symmetries, and that an extension such as the Tsallis entropy~\cite{Tsallis:2012js} is better suited for our solutions.

Plugging in the expression for the outer horizon in Eq.~(\ref{eq:horcase1}), (where the Hawking temperature is canonically defined) we find (working with $k<0$ from now on)
\begin{equation}
    \begin{aligned}
    T_{H,1} &= -\frac{1}{4\pi k} +\mathcal{O}((s^{00})^2) \\
    T_{H,{\rm trace}} &= -\frac{1}{4\pi k}+\frac{1}{8\pi k}s^{00}+\mathcal{O}((s^{00})^2)
    \end{aligned}
\end{equation}

where we find that for the inner horizon, a consistently divergent Hawking temperatures arises for both cases, and we will not consider this further.\footnote{If we were to include terms quadratic terms in $s^{00}$ (this is beyond our approximation), we would find that $T_{H,1}$ is given a correction quadratic in $s^{00}$, whereas the correction is linear for $T_{H,\rm trace}$.}

In order to obtain the entropy of these new solutions, we notice that in modified gravity, especially models featuring higher derivatives or broken spacetime symmetries, the GR notion of proportionality of horizon entropy to horizon area generally does not hold. We will instead obtain the entropy by assuming explicitly that the first law of thermodynamics $dk=T_H dS$ stays valid\footnote{Here, we use the mass parameter $k$ which is clearly negative, and the thermodynamic quantities will therefore also be shifted by a sign.}. Other approaches have been considered in the literature, for example that of extended thermodynamics of bumblebee black holes \cite{Mai:2023ggs}, which we do not consider here.

Using the definition for $T_H$ and the expressions for the outer horizon $r_+$ in terms of the mass parameter $k$ we can derive the entropy as
\begin{equation}
    S := \int \frac{dk}{T_H(k)}+S_0 = \int dr_+ \frac{1}{T_H} \frac{dk}{dr_+} + S_0,
\end{equation}

and we obtain to linear order, by explicit integration 
\begin{equation}
    \begin{aligned}
        S_1 =& -\frac{4\pi r_+^3}{3|k|}+\frac{\pi(r_+^3-3kr^2)}{3|k|}s^{00}+S_0, \\
        S_{1,\rm{trace}} =& -\frac{4\pi r_+^3}{3|k|}+\frac{\pi(2r_+^3-3kr^2)}{3|k|}s^{00}+S_0,
    \end{aligned}
\end{equation}
where we have expanded in $s^{00}$ in order to find a well-behaved solution. The two are identical except for a factor of two in the second term. Also, when solving for the mass parameter $k$ we obtain two solutions, both of which lead to the same entropy at linear order. The integration constant $S_0$ must be fixed by some physical consideration, such as the counting of degrees of freedom or the existence of a smooth limit when the horizon area vanishes. We note also that the common logarithmic corrections to black-hole entropy appears as $\ln{(s^{00}-2)}(s^{00})^2$ in both $S_1$ and $S_{1,\rm{trace}}$, but lies beyond linear order.

\subsubsection*{Spacetime singularities}\label{sec:case1sing}
In order to make the singularities explicit, we compute the standard curvature scalars for Case 1 by means of a series expansion, which read
\begin{equation}\label{eq:case1scalars}
    R_1=0, \quad K_1=R_{\mu\nu\alpha\beta}R^{\mu\nu\alpha\beta} = \frac{12k^2}{r^6} - \frac{12k^3s^{00}}{r^7},
\end{equation}
where we note that the Ricci scalar is zero to first order in $s^{00}$, whereas the Kretchmann scalar $K_1$ takes on non-zero values. This expression reveals no curvature singularities at the horizons\footnote{Should we include higher-order terms in $K_1$, singularities at the new horizons would appear. However, this lies beyond our linear approximation.},
but an interesting feature appears, which can be seen in Figure~\ref{fig:kl1}. Due to the $k^3$-dependence in the extra term of $K_1$, there exists an equilibrium point at $r=ks^{00}$, beyond which the $r^{-7}$ term in (\ref{eq:case1scalars}) dominates. Therefore, when $s^{00}<0$, $K_1$ exhibits a negative divergence as $r\to0$. For positive $s^{00}$, the behaviour of $K_1$ is similar to Schwarzschild except that it is $\propto r^{-7}$ rather than $r^{-6}$.

\begin{center}
    \begin{figure}[h!]
    \centering
        \includegraphics[scale=0.7]{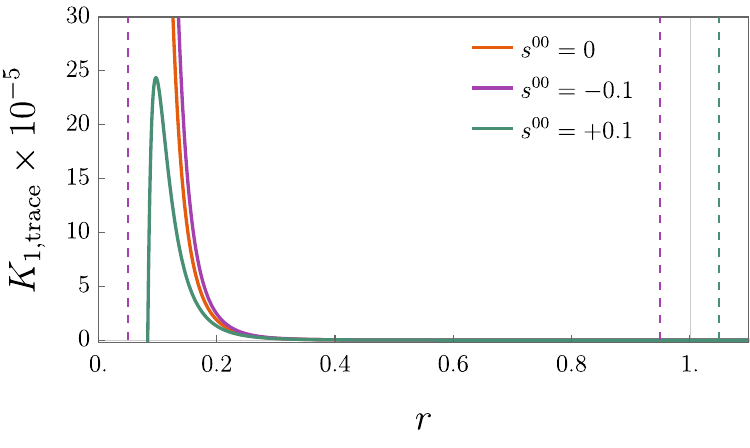}
        \caption{The Kretschmann scalar for the trace subset of Case 1, where the $s^{00}=0$ case is that of GR and Schwarzschild. Dashed lines indicate the horizons for the corresponding solution, and the solid grey line represents the Schwarzschild horizon.}
        \label{fig:kl1trace}
    \end{figure}
\end{center}

In the case of the trace-term solution in Eq.~(\ref{eq:tracesol}), we find that both the Ricci and Kretschmann scalars take on a linear correction as
\begin{equation}\label{eq:Kretschtrace}
\begin{aligned}
       R_{1, {\rm trace}} =& -3\frac{k^2}{r^4}s^{00} \\
   K_{\rm1, trace} =& \frac{12k^2}{r^6} + \frac{12s^{00}\left(k^3+2k^2r\right)}{r^7},
\end{aligned}
\end{equation}
which shows similar behaviour compared to the previous solution considered. However, we note that the situation is now reversed, and the negative divergence in both the Ricci and Kretschmann scalars now occur for opposite signs of $s^{00}$, as can be read off from Eq.~(\ref{eq:Kretschtrace}) as well as seen in Figure~\ref{fig:kl1trace}.

\subsection{Case 2}\label{sec:propcase2}

\subsubsection*{Horizons}\label{sec:case2hor}

For the case of $s^{11}\neq0$ in Eq.~(\ref{eq:scase2}) we find that the horizons must obey
\begin{equation}
    \left(1-s^{11}\right)r_\star^2+k\left(2-s^{11}\right)r_\star+k^2\left(1+\tfrac{3}{4}s^{11}\right) = 0,
\end{equation}
the solution to which can be written as
\begin{equation}\label{eq:rstar2}
    r_{\star, 2} = \frac{k(2-s^{11})}{2(s^{11}-1)}\pm \frac{|k|}{2(s^{11}-1)}\sqrt{s^{11}(4s^{11}-3)},
\end{equation}
where the subscript $``2"$ refers to Case 2, and which to linear order can be taken as
\begin{equation}
\label{eq:case2hor}
r_{\star, 2} = -k \left(1 + \frac{s^{11}}{2} \pm \sqrt{-\frac{3}{4} s^{11}} \right) + \mathcal{O}(( s^{11})^2)
\end{equation}
When $s^{11}<0$, two real solutions for $r_\star$ exist; however, when $s^{11}>0$, both solutions are imaginary and we are left with a naked singularity. These solutions do not coincide with Killing horizons of this Case, as is shown in Appendix~\ref{app:killinghorizons}.

\subsubsection*{Thermodynamics}\label{sec:case2thermo}
\begin{center}
    \begin{figure}[h!]
    \centering
        \includegraphics[scale=0.7]{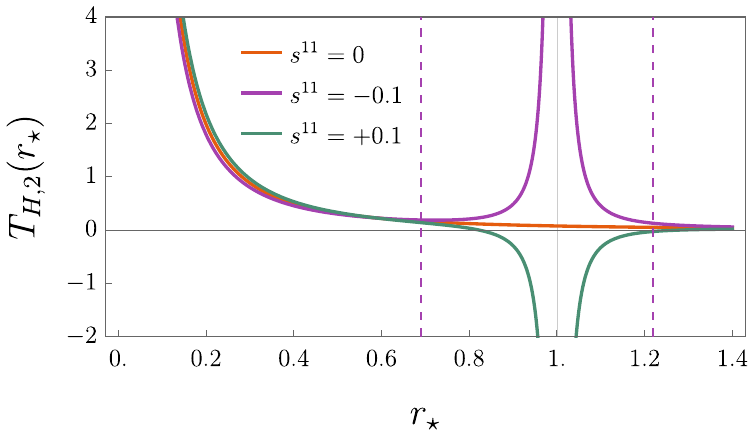}
        \caption{The Hawking temperature for Case 2 as a function of the horizon radius $r_\star$, where the $s^{00}=0$ case is that of GR and Schwarzschild. Dashed lines indicate the horizons for the corresponding solution, and the solid grey line represents the Schwarzschild horizon.}
        \label{fig:TH2}
    \end{figure}
\end{center}

Due to the non-standard horizon structure, we may expect non-standard behaviour in the thermodynamic variables of such horizons. First, we form the Hawking temperature according to Eq.~(\ref{eq:hawkingdef}) and find (in terms of the horizon radius $r_\star$)
\begin{equation}
    T_{H,2} = \frac{|k|}{4\pi r_\star^2}+\frac{|k|(r_\star^2+4kr_\star+2k^2)}{8\pi r_\star^2(k+r_\star)^2}s^{11}
\end{equation}
which in the GR limit is a monotonically decaying function of $r_\star$, as can be seen in Figure~\ref{fig:TH2}. Plugging in the expression for the horizons (\ref{eq:rstar2}) for negative $s^{11}$ reveals a monotonically {\it increasing} Hawking temperature, but due to the linear nature of the metric function we find a rescaled GR limit when $s^{11}\to0$. 
Allowing $s^{11}$ to be non-zero introduces divergencies in the Hawking temperature at the Schwarzschild horizon, which is not a horizon when $s^{11} \neq 0$; no temperature divergencies appear at the new horizons in this solution. Furthermore, forming the thermodynamic variables we find
\begin{equation}
    \begin{aligned}
        T_{H,2} =& \frac{5}{12\pi|k|}-\frac{5\sqrt{-s^{11}}}{12\sqrt{3}\pi k}-\frac{35s^{11}}{48\pi|k|}, \\
        S_2 =& \frac{288\pi k^3\big(2-s^{11}\pm\sqrt{s^{11}(4s^{11}-3)}\big)r}{5|k|(4+3s^{11}(4\sqrt{3}|k|\sqrt{-s^{11}}+3k(7s^{11}-4)))}
    \end{aligned}
\end{equation}
which satisfies the first law to first order in $s^{11}$ and is real for $s^{11}<0$ which constrains $s^{11}$ to be negative. $S_2$ can of course be expanded to linear order again, after which we obtain $S_2=-\frac{12\pi |k|}{5}r_+ - \frac{2\pi |k|}{5k}(kr_++3|k|r_+)s^{11}-\frac{2\pi\sqrt{3}|k|}{5k}(3kr_+\sqrt{-s^{11}}+2|k|r_+\sqrt{s^{11}})+S_0$.

The divergence in the Hawking temperature in Fig.~\ref{fig:TH2} is reminiscent of a first-order phase transition of Hawking-Page type \cite{Hawking:1982dh}\footnote{Although the Hawking-Page transition is characterised through the heat capacity.} between a large stable black hole and a thermal gas in AdS space. As in the case of negative Hawking temperature in Case 1, this may also signify that extended thermodynamics is necessary; however, since we also observe {\it curvature} singularities (see Figures~\ref{fig:ricci2}, \ref{fig:k2}), we may conclude that this solution has a true singularity between the two horizons. Since the outer horizon is not a true event horizon, but rather a throat, this new singularity is not future-pointing, and the interior spacetime is inaccessible due to the presence of this spherical singularity, akin to the black-hole {\it firewall}~\cite{Raju:2020smc}.

\subsubsection*{Spacetime singularities}\label{sec:case2sing}
\begin{figure}
    \centering
    \begin{subfigure}[T]{0.49\textwidth}
        \includegraphics[scale=0.6]{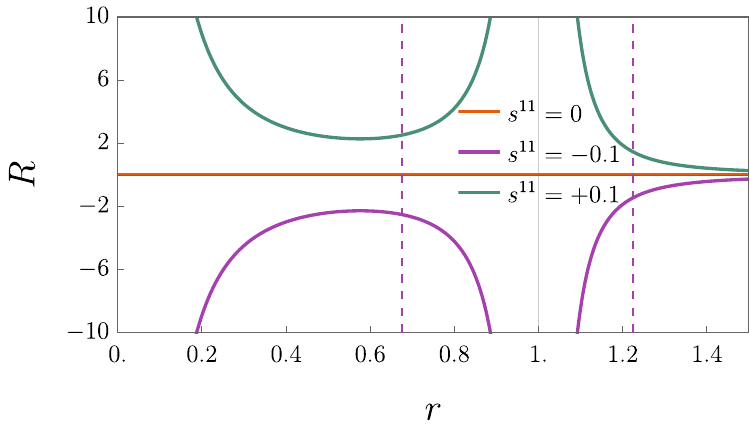}
        \caption{The Ricci scalar for Case 2, where the $s^{00}=0$ case is that of GR and Schwarzschild. Dashed lines indicate the horizons for the corresponding solution, and the solid grey line represents the Schwarzschild horizon.}
        \label{fig:ricci2}
    \end{subfigure}
    \hfill
        \begin{subfigure}[T]{0.49\textwidth}
        \includegraphics[scale=0.6]{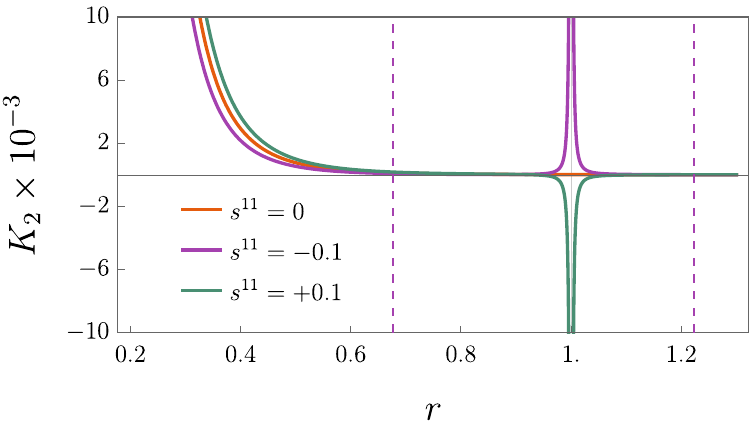}
        \caption{The Kretschmann scalar for Case 2, where the $s^{00}=0$ case is that of GR and Schwarzschild. Dashed lines indicate the horizons for the corresponding solution, and the solid grey line represents the Schwarzschild horizon.}
        \label{fig:k2}
    \end{subfigure}
    \caption{Ricci scalar (left) and Kretchmann scalar (right) for Case 2.}
    \label{fig:case2curvature}
\end{figure}

As in the previous cases, we compute the relevant curvature scalars for the solution (\ref{eq:scase2}), and we find that both the Ricci and Kretschmann scalars show non-trivial behaviour linear in $s^{11}$ as
\begin{equation}\label{eq:Riccicase2}
    \begin{aligned}
    R_2 =& \frac{\left(3k^2+4kr+2r^2\right)}{r^2\left(k+r\right)^2}s^{11}, \\
    K_2 =& \frac{12k^2}{r^6} + \frac{12k\left(9k^3+6k^2r-8k^2r^2-4r^3\right)}{r^6(k+r)^2}s^{11}.
    \end{aligned}
\end{equation}
The non-zero Ricci scalar represents a significant difference from GR, and even from the solutions found for Case 1. As can be seen in Figure~\ref{fig:ricci2}, no horizons appear for $s^{11}>0$, and both singularities are therefore naked, where we notice that the manifold is {\it not} Ricci flat outside the horizon except asymptotically.\footnote{Therefore the linearised-gravity constraint $s^{\mu\nu}\partial_\mu R_L=0$~\cite{ONeal-Ault:2020ebv} must be understood in the asymptotic sense.} We also examine the Kretschmann scalar in the same way in Figure~\ref{fig:k2}, where the extra singularity at the Schwarzschild horizon (which is naked for $s^{11}>0$) appears. Inside the second horizon, the solution diverges similarly to Schwarzschild.

\subsection{Case 3}\label{sec:propcase3}

\subsubsection*{Horizons}\label{sec:case3or}
The case when both $s^{00}$ and $s^{11}$ are non-zero can be obtained as in Eq.~(\ref{eq:scase3}). In this case we have two free parameters, and the horizons are determined by
\begin{equation}\label{eq:horcase3}
    \begin{aligned}
        \left(1-s^{11}\right)r_\star^3 + k\left(2-s^{11}\right)r_\star^2 + &k^2\left(1-\tfrac{1}{4}s^{00}+\tfrac{3}{4}s^{11}\right)r_\star \\&-\tfrac{k^3}{4}s^{00} = 0,
    \end{aligned}
\end{equation}
which has three real solutions for and $s^{11}<0$ (only two physical horizons appear for $s^{00}<0$) which to first order read
\begin{equation}\label{eq:case3horizonsLinear}
    \begin{aligned}
    r_{\star, 1 2} =& -k \left(1 \pm \frac{1}{2} \sqrt{-3s^{11}} + \frac{s^{00}}{8} + \frac{s^{11}}{2} \right), \\
    r_{\star,3} =& \frac{k}{4}s^{00}.
    \end{aligned}
\end{equation}
It is worth pointing out that the above solutions do not have smooth limits to the cases considered above, where only one of $\left\{s^{00},s^{11}\right\}$ are non-zero. Once the solution $r_{\star,3}$ has been found, we can factorise Eq.~(\ref{eq:horcase3}) and identify a quadratic polynomial with discriminant $\Delta$, and the ``discontinuity'' arising when $s^{11} \to 0$ may be an artefact from truncating $\Delta$ to first order in the SME coefficients. For correspondence between the different cases considered here, we would need to include second-order terms in $\Delta$; in fact, it is easily checked that the appearance of an $(s^{00})^2$-term guarantees such a match, but this lies beyond our approximation. Similar to Case 2, the Killing horizons do not coincide with these horizons, as is shown in Appendix~\ref{app:killinghorizons}. We note that the $s^{11}$-terms are responsible for this effect, since Case 1 is homogeneous at linear order in the coefficients, whereas Case 2 is not.

\subsubsection*{Thermodynamics}\label{sec:case3hermo}
\begin{figure}
    \centering
    \begin{subfigure}[T]{0.48\textwidth}
        \includegraphics[width=\textwidth]{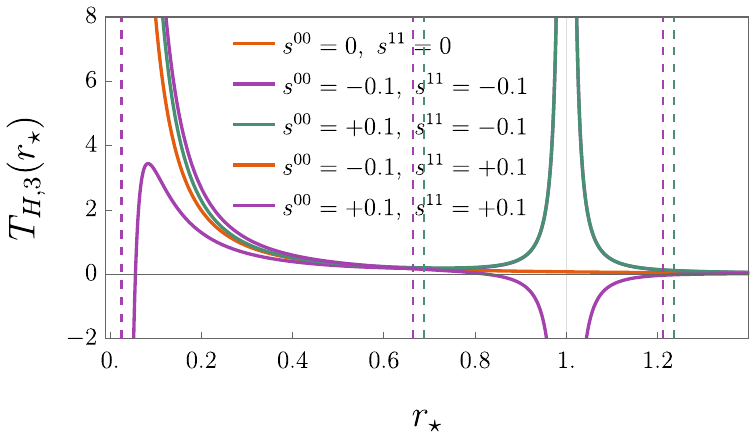}
        \caption{The Hawking temperature for Case 3 as a function of the horizon radius $r_\star$, where the $s^{00}=s^{11}=0$ case is that of GR and Schwarzschild. Dashed lines indicate the horizons for the corresponding solution, and the solid grey line represents the Schwarzschild horizon. Imaginary or negative horizons have not been included.}
        \label{fig:TH3}
    \end{subfigure}
    \hfill
        \begin{subfigure}[T]{0.48\textwidth}
        \includegraphics[width=\textwidth]{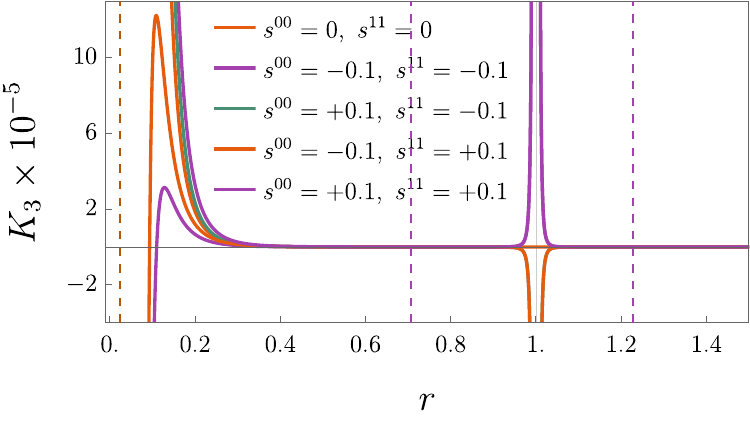}
        \caption{The Kretschmann scalar for Case 3, where the $s^{00}=s^{11}=0$ case is that of GR and Schwarzschild. Dashed lines indicate the horizons for the corresponding solution, and the solid grey line represents the Schwarzschild horizon.}
        \label{fig:K3}
    \end{subfigure}
    \caption{Hawking temperature (left) and Kretchmann scalar (right) for Case 2.}
    \label{fig:case3curvature}
\end{figure}

The Hawking temperature reads
\begin{equation}
    T_{H,3} = \frac{|k|}{4\pi r_\star^2}-\frac{k|k|}{8\pi r_\star^3}s^{00}+\frac{|k|(r^2+4kr_\star+2k^2)}{8\pi r_\star^2(k+r_\star)^2}s^{11}.
\end{equation}
By plugging in the three horizons in (\ref{eq:case3horizonsLinear}), we find that $r_{\star, 12}$ both have smooth limits when $\{s^{00},s^{11}\}\to 0$ simultaneously, and when $s^{11}\to0$ on its own, but is singular for $s^{00}\to 0$; $r_{\star,3}$ is singular for when $s^{00}\to0$ on its own. The full order expressions for $r_{\star,12}$ are lengthy, and we do not display them here. The linearised Hawking temperature for this case can be seen in Figure~\ref{fig:TH3},
where we see that only $\{s^{00}=-0.1,s^{11}=-0.1\}$ and $\{s^{00}=+0.1,s^{11}=-0.1\}$ have at least one positive and real horizon. As in the previous case, turning on spacetime-symmetry breaking causes divergent behaviour in the Hawking temperature at the Schwarzschild horizon, and certain parameter ranges results in a negative divergence close to $r_\star=0$. 

We compute the entropy of all three cases listed in Eq.~(\ref{eq:case3horizonsLinear}) (which can be easily solved for $k$), and we find that the entropy related to $r_{\star, 1}$ and $r_{\star,2}$ is the same and reads
\begin{equation}
    \begin{aligned}
        S_{3,12} =& -\frac{4\pi r^3}{3|k|}+\frac{2\pi r^3}{3|k|}\sqrt{-s^{11}}+\frac{\pi r^2(r-6k)}{6|k|}s^{00} \\&+\frac{\pi}{3|k|(r+k)}\Big[6k^4+(k+r)\big(24k^3\ln{\left(1+\frac{r}{k}\right)} \\&+r(7r^2+6kr-8k^2)\big)\Big]s^{11},
    \end{aligned}
\end{equation}
and for $r_{\star, 3}$, we find a singularity when $s^{00}\to 0$; we do not consider this case further.

\subsubsection*{Spacetime singularities}\label{sec:case3sing}
Computing the singularities to linear order in the same manner as in previous sections, we find the following expression for the Ricci scalar
\begin{equation}
    R_3 = \frac{2r^2+4kr+3k^2}{r^2(k+r)^2},
\end{equation}
which is {\it independent} of $s^{00}$, revealing another discontinuity. This is because the following terms in the expansion of the Ricci scalar are of order $s^{00}s^{11}$ and higher, and we treat them as higher-order terms; again, we note that quadratic corrections are necessary for a proper GR limit here. Thanks to the linear nature of the metric solutions, we obtain exactly the same solution as for Case 2 in Eq.~(\ref{eq:Riccicase2}). For the Kretschmann scalar, we obtain
\begin{equation}
    \begin{aligned}
        K_3 =& \frac{12k^2}{r^6}-\frac{12k^3}{r^7}s^{00}-\frac{2k(4r^3+8kr^2-6k^2r-9k^3)}{r^6(k+r)^2}s^{11},
    \end{aligned}
\end{equation}
the behaviour of which can be seen in Figure~\ref{fig:K3}.
We notice here that the features of the Kretschmann scalar is similar to that of Case 1+Case 2, again due to the linear nature of the solutions.

\section{Photon radial geodesics}\label{sec:photonradialgeodesics}
We move now to phenomenological aspects of these new solutions, starting with photon radial geodesics (restoring the value of $c$ temporarily). In this case, we can write the metric as
\begin{equation}
    ds^2 = -c^2N^2(r)dt^2+\frac{1}{f(r)}dr^2,
\end{equation}
and since we have $ds^2=0$ for null rays, we get
\begin{equation}\label{eq:nullradial}
    cdt = \pm \frac{dr}{\sqrt{N^2f}}.
\end{equation}
If we now expand $N$ and $f$ as in Eq.~(\ref{eq:eqdiff1})-(\ref{eq:solpart1}), $X \approx X_0 + X_1$, and we can write as a linear approximation in any small parameter $\left( N^2f \right)^{-\frac{1}{2}} \approx \left( N_0^2 f_0 \right)^{-\frac{1}{2}} \left( 1 - N_1 / N_0 - f_{1}/(2f_0) \right)$; we use this to compute the radial null geodesics for all cases below.

\subsection{Case 1}
For the Case 1 solution in Eq.~(\ref{eq:scase1}), we have
\begin{equation}
    \frac{1}{\sqrt{N^2f}} = \left(1+\frac{k}{r}\right)^{-1}+s^{00}\frac{k^2}{4}\frac{1}{(k+r)^2} + \mathcal{O}((s^{00})^2), 
\end{equation}
which we use to integrate Eq.~(\ref{eq:nullradial}) to
\begin{equation} \label{eq:rayeq1}
    ct = \pm\left[r-k\ln{\left( 1 + \frac{r}{k} \right)}-\frac{s^{00}}{4}\frac{k^2}{r+k}\right] + A,
\end{equation}

where $A$ is an integration constant which depends on the asymptotic conditions of the null ray. This exhibits exotic behaviour close to the Schwarzschild horizon for $s^{00}<0$ but recovers Minkowski lightcones asymptotically.
Similarly, we look at the trace-term case and find
\begin{equation} \label{eq:rayeq2}
    \frac{1}{\sqrt{N^2f}}\approx \frac{r}{k+r}-\frac{k(k+2r)}{2(k+r)^2}s^{00},
\end{equation}
which we integrate to obtain
\begin{equation}
\label{eq:radgeos00}
    ct=\pm \left[ r-k \left( 1 + s^{00} \right) \ln{\left( 1 + \frac{r}{k} \right)}- \frac{s^{00}}{2} \frac{k^2}{k + r} \right] + B,
\end{equation}
which shows behaviour similar to that of Case 1 above.

\subsection{Case 2}
For the Case 2 solution in Eq.~(\ref{eq:scase2}) we have that
\begin{equation}
\frac{1}{\sqrt{N^2f}}\approx  \left(1+\frac{k}{r}\right)^{-1}+\frac{r(r^2-2k^2)}{2(k+r)^3}s^{11},
\end{equation}
which can be integrated to find 

\begin{equation}
\label{eq:radgeos11}
    \begin{aligned}
        ct &= \pm \biggl[ \left(1 + \frac{s^{11}}{2} \right) r - \left(1 + \frac{3 s^{11}}{2} \right) k \ln{\left( 1 + \frac{r}{k} \right)} \\&- \frac{s^{11}}{2} \left( \frac{k^2}{r+k} + \frac{1}{2} \frac{k^2}{\left( r+k \right)^2} \right) \biggr] + C.
    \end{aligned}
\end{equation} 
We remark that we do not recover the Minkowski lightcone asymptotically; indeed, contrary to Eq.~(\ref{eq:rayeq1}) or (\ref{eq:rayeq2}), $ct/r \to 1 + \frac{s^{11}}{2}$ at spatial infinity for this case. As such, this solution leaves a characteristic imprint of diffeomorphism breaking at spatial infinity in the form of an angular deficit, as in the case of cosmic strings and other topological defects.

\subsection{Case 3}
For the solution in Eq.~(\ref{eq:scase3}), we have that
\begin{equation}
    \frac{1}{\sqrt{N^2f}}\approx \left(1+\frac{k}{r}\right)^{-1} + s^{00} \frac{k^2}{4} \frac{1}{\left(k+r \right)^2} + s^{11}\frac{r(r^2-2k^2)}{2(k+r)^3},
\end{equation}
and integrating this expression, we obtain

\begin{equation}
\label{eq:radgeoall}
    \begin{aligned}
        ct &= \pm\Big[\left(1+\frac{s^{11}}{2}\right)r-\left(1+\frac{3s^{11}}{2}\right)k\ln{\left( 1 + \frac{r}{k} \right)} \\&-\frac{s^{11}}{2}\Big(\frac{k^2}{r+k}+\frac{k}{2(r+k)^2}\Big) -\frac{s^{00}}{4}\frac{k^2}{r+k}\Big] + D.
    \end{aligned}
\end{equation}

Since the process is entirely linear, we find that the radial geodesics (\ref{eq:radgeoall}) are the sum of Eqs.~(\ref{eq:radgeos00}) and (\ref{eq:radgeos11}); 
it is clear from this expression that the dynamics close to the Schwarzschild horizon are governed by $s^{11}$ rather than $s^{00}$.

\section{Conserved quantities and orbits}\label{sec:consquantsorbits}
In order to derive classical observables such as periastron precession, we must solve the general geodesic equation in our new spacetime. We denote the four-velocity of a test particle by $u^\mu = dx^\mu/d\tau$, where $\tau$ is an affine parameter on the geodesic and $u^\mu$ is normalised such that $u_\mu u^\mu = \pm 1$ or $0$, depending on the nature of the curve (spacelike, timelike or null). For a general $N^2f$ spacetime we have that
\begin{equation}
    \begin{aligned}
        g_{\mu\nu}u^\mu u^\nu =& -N^2\left(\frac{dt}{d\tau}\right)^2 +\frac{1}{f}\left(\frac{dr}{d\tau}\right)^2+r^2\left(\frac{d\theta}{d\tau}\right)^2  \\&+r^2\sin{\theta} \left(\frac{d\phi}{d\tau}\right)^2 \\ =& -\epsilon,
    \end{aligned}
\end{equation}
where $\epsilon = \{\pm 1, 0\}$ represents spacelike, timelike, and null geodesics, respectively. We also identify the Killing vector fields as $K_t^\mu = (\partial_t)^\mu$ and $K_\phi^\mu = (\partial_\phi)^\mu$ for a static and spherically symmetric spacetime, and we identify the associated conserved quantities as energy ($E$) and angular momentum ($L$) as
\begin{equation}
\label{consttraj}
    \begin{aligned}
    E \equiv& K_t u = -N^2\frac{dt}{d\tau}, \\
    L \equiv& K_\phi u = r^2\sin{\theta} \frac{d\phi}{d\tau}.
    \end{aligned}
\end{equation}
We note that as $E$ and $L$ depend on the geodesic, and we consider here GR geodesics perturbed by $s^{00}$, the conserved quantities can be decomposed as
$E = E_0 + s^{00} E_1 + \mathcal{O}(( s^{00})^2), L = L_0 + s^{00} L_1 + \mathcal{O}((s^{00})^2)$, which we will use when necessary below. 

Using these variables, the geodesic equation can be written as
\begin{equation}
\label{eq:energ1}
    -\frac{E^2}{N^2}+\frac{1}{f}\left(\frac{dr}{d\tau}\right)^2+\frac{L^2}{r^2} = -\epsilon,
\end{equation}
and we can specify $\theta=\pi/2$ without loss of generality. 

\begin{figure}
    \centering
    \begin{subfigure}[T]{0.49\textwidth}
        \includegraphics[width=\textwidth]{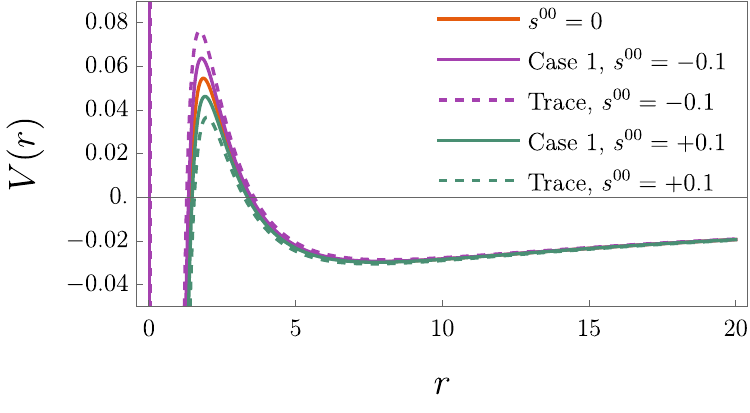}
        \caption{The potential $V(r)$ in the case of massive particles for Case 1 (solid) and the trace case (dashed), where $s^{00}=0$ is that of GR and Schwarzschild for the same value of $L=2.1$; in the trace case, $E=0.2$ for illustrative purposes.}
        \label{fig:potentialcase1andtrace}
    \end{subfigure}
    \hfill
        \begin{subfigure}[T]{0.49\textwidth}
        \includegraphics[width=\textwidth]{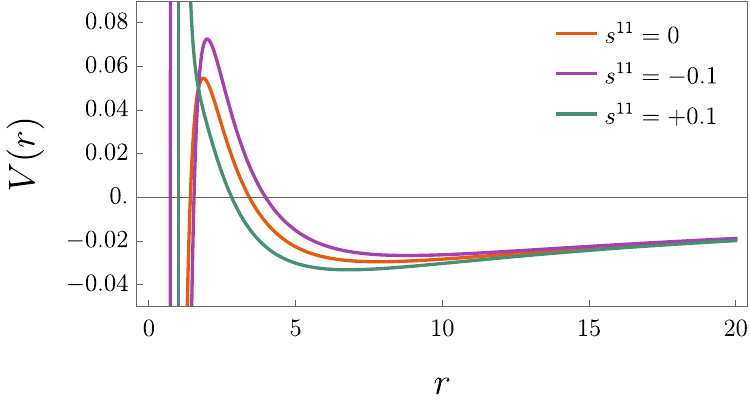}
        \caption{The potential $V(r)$ in the case of massive particles for Case 2, where $s^{11}=0$ is that of GR and Schwarzschild for the same value of $L=2.1$; in the trace case, $E=0.2$ for illustrative purposes.}
        \label{fig:potentialcase2}
    \end{subfigure}
    \caption{Effective potential for massive particles.}
\end{figure}

We start with the simplest situation, which is Case 1, where we only have $s^{00}\neq 0$. In this case, the above geodesic equation can be written as
\begin{equation}
\label{eq:energycons}
\begin{aligned}
    \frac{1}{2}(E^2-\epsilon) &= \frac{1}{2}\left(\frac{dr}{d\tau}\right)^2+ \frac{\epsilon k}{2r}+\frac{kL^2}{2r^3} -s^{00}\frac{k^2L^2}{8r^4}\\&+\frac{1}{2r^2}\left(L^2-s^{00}\frac{\epsilon k^2}{4}\right).
\end{aligned}
\end{equation}
In the above equation, we identify three different terms that we name "mechanical energy" ($E_m$), "kinetic energy" ($T$) and "potential energy" ($V$) for their resemblance with their Newtonian counterparts (resetting units to $c=1$ from now on)
\begin{equation}\label{eq:TandVcase1}
    \begin{aligned}
        E_m =& \frac{1}{2}(E^2-\epsilon), \\
        T =& \frac{1}{2}\left(\frac{dr}{d\tau}\right)^2, \\
        V(r) =& \frac{\epsilon k}{2r}+\frac{1}{2r^2}\left(L^2-s^{00}\frac{\epsilon k^2}{4}\right) + \frac{kL^2}{2r^3}-s^{00}\frac{k^2L^2}{8r^4},
    \end{aligned}
\end{equation}
where the shape of the potential $V(r)$ determines the allowed orbits around such a body; we note that $dV/dr=0$ is a necessary and sufficient condition for circular orbits\footnote{However, this is not enough to determine whether they are stable.} Focusing on the orbits of massive particles ($\epsilon=1$) from now on, we notice that $dV/dr$ is a fifth-order polynomial, which we may solve analytically; however, these solutions are very messy and give us little intuition as to the circular orbits. Instead, we use a perturbative ansatz of the form $\widetilde{r}_{\rm sch} = r_{\rm sch}+ \xi \, s^{00}$, where $\xi$ is a constant with dimensions of mass and $r_{\rm sch}$ is the solution to $dV/dr=0$ in the Schwarzschild case. Using such an approximation we can isolate the shift of the circular orbits away from $r_{\rm sch}$ to the new radius $\widetilde{r}_{\rm sch}$ due to terms linear in $s^{00}$, since the Schwarzschild case is decoupled in the expression for $dV/dr$ and thus equates to zero separately. The perturbative ansatz can be written as
\begin{equation}
\label{eq:dvlim}
   \widetilde{r}_{\rm sch}^{-n} \approx r_{\rm sch}^{-n}-\xi s^{00}\frac{n}{r_{\rm sch}^{n+1}},
\end{equation}
and after some algebra, we obtain

\begin{equation}\label{eq:dVdrcase1}
    \begin{aligned}
       \frac{dV}{dr}\Big|_{r\to\widetilde{r}_{\rm sch}} =& s^{00}\Bigg[\frac{4 k \xi - \left( 8 L_0 L_1 - k^2 \right)  }{4r_{\rm sch}^3}\\&+\frac{3L_0 \left(\xi L_0 - k L_1 \right)}{r_{\rm sch}^4}\\&+\frac{kL_0^2(12\xi+k)}{2r_{\rm sch}^5}\Bigg] = 0,
    \end{aligned}
\end{equation}
where we can solve for $\xi$ as $\xi= ( [8 L_0 L_1 - k^2 ] r_{\rm sch}^2 + 12 k L_0 L_1 r_{\rm sch} - 2k L_0^2 )/[4 ( k r_{\rm sch}^2 + 3 L_0^2 r_{\rm sch} + 6 k L_0^2 )  ]  $, which solves the system. We note that the denominator vanishes for $L_0^2 = -r_{\rm sch}^2/3$ and $\xi$ is therefore finite for real values of $L_0$. For $s^{00}\sim \pm 0.1$, two circular orbits still exist, with their locations shifting around $1\%$\footnote{This difference is much smaller for the case of massless particles.}, although the height of the potential shifts on the order of $10\%$, altering the angular momentum required to reach a circular orbit. Interestingly, for $s^{00}<0$, a potential barrier appears close to $r=0$ where the last term in Eq.~(\ref{eq:dVdrcase1}) dominates, as can be seen in Figure~\ref{fig:potentialcase1andtrace}. 

For the trace case, which we have shown to be similar to Case 1, we have the same mechanical energy $E_m$ as above; however, the potential has an important difference, and reads
\begin{equation}
    \begin{aligned}
        V(r) =& \frac{L^2}{2r^2}\left(1+\frac{k}{r}-s^{00}\frac{k^2}{2r^2}\right)\\&+\frac{\epsilon k}{2r} +\frac{s^{00}k}{r}\left(E^2-\frac{\epsilon k}{4r}\right),
    \end{aligned}
\end{equation}
where a non-standard term proportional to $E^2$ appears. In Figure~\ref{fig:potentialcase1andtrace} we show the potential for Case 1 and the trace case together for the same value of $L$; it is clear from this plot that $E$ plays the role of symmetrically increasing the amplitude of the potential around the Schwarzschild case. Altering the value of $L$ acts as a scaling, as expected, and we note that the same potential barrier appears for the trace case when $s^{00}<0$.

We obtain the mechanical, kinetic and potential energy for Case 2 in the same way, leading to
\begin{equation}
\label{energyphotcase2}
    \begin{aligned}
        E_m =& \frac{1}{2}(E^2-\epsilon)(1-s^{11}), \\
        T =& \frac{1}{2}\left(\frac{dr}{d\tau}\right)^2, \\ 
        V(r) =& \frac{k}{2r}+\frac{L^2}{2r^2}(1-s^{11})+\frac{kL^2}{2r^3} \\&+s^{11}\left(1+\frac{k}{r}\right)^{-1}\Big[\frac{3\epsilon k^2}{8r^2}+\frac{3k^2L^2}{8r^4} \\&-\frac{E^2k^2}{4r^2}\left(1+\frac{k}{r}\right)^{-1}\Big],
    \end{aligned}
\end{equation}
where we note a scaling of the mechanical energy $E_m$ as well as a coupling between the potential $V$ and the conserved quantity $E$ associated with energy. We also note the existence of terms which diverge at $r\to -k$ (depending on the sign of $s^{11}$). For $s^{11}>0$, an infinite barrier exists at the Schwarzschild horizon, and crossing into the interior is not possible for a massive particle -- this can also be seen from the behaviour of the curvature invariants in Figure~\ref{fig:case2curvature} and \ref{fig:case3curvature}. Despite this, both stable and unstable circular orbits exist outside this horizon for $s^{11}>0$, whereas only a stable orbit exists around $r=5$ for $s^{11}<0$, as can be seen in Figure~\ref{fig:potentialcase2}.
Finally, for Case 3, we find as sum of the effects in Case 1 and 2, where the mechanical energy $E_m$ is the same as in Case 2, and the potentials reads
\begin{equation}
    \begin{aligned}
        V(r) =& \frac{k\epsilon}{2r}+\frac{kL^2}{2r^3}-s^{00}\frac{k^2L^2}{8r^4} +\frac{1}{2r^2}\left(L^2(1-s^{11})-s^{00}\frac{k^2}{4}\right) \\& +s^{11}\left(1+\frac{k}{r}\right)^{-1}\Big[\frac{3\epsilon k^2}{8r^2}+\frac{3k^2L^2}{8r^4} -\frac{E^2k^2}{4r^2}\left(1+\frac{k}{r}\right)^{-1}\Big],
    \end{aligned}
\end{equation}
where we observe the same divergent behaviour as in Case 2 and a similar potential barrier close to $r=0$ as in Case 1.

\section{Periastron precession}\label{sec:precession}
In order to study the precession of the periastron, we investigate the angular motion in the equatorial plane. From the previous Section, we know that the geodesic equation can be written in the form
\begin{equation}
    E_m = \frac{1}{2}\left(\frac{dr}{d\tau}\right)^2 + V(r). 
\end{equation}
We introduce the new variable $u := 1/r$, through which we can write
\begin{equation}
    E_m = \frac{L^2}{2}\left(\frac{du}{d\phi}\right)^2 + \widetilde{V}(u),
\end{equation}
where we have made use of the relations $dr/d\tau = dr/d\phi \times d\phi/d\tau$ and $dr/d\phi \times L/r^2 = -L du/d\phi$. Taking a derivative, we then obtain
\begin{equation}\label{eq:precessioneq}
    L^2\frac{d^2u}{d\phi^2} + \widetilde{V}'(u) = 0,
\end{equation}
which is the relation necessary for studying the precession. 

\subsection*{Almost circular orbits}

We focus now on Case 1 and leave the rest for future study: starting by considering approximately circular orbits, we introduce the expansion
\begin{equation}\label{eq:uexp}
    u = u_c (1+w(\phi)), \quad w(\phi) \ll 1,
\end{equation}
where $u= u_c$ corresponds to the exactly circular case. For Case 1, the potential $V(r)$ was calculated in Eq.~(\ref{eq:TandVcase1}) (second equation), from which we find
\begin{equation}
        \begin{aligned}
            \widetilde{V}(u) = \frac{k}{2}u& + \frac{1}{2}\left(L^2-s^{00}\frac{k^2}{4}\right)u^2 + \frac{kL^2}{2}u^3 - s^{00}\frac{k^2L^2}{8}u^4.
        \end{aligned}
\end{equation}
We now expand $\widetilde{V}(u)$ using (\ref{eq:uexp}) and obtain Eq.~(\ref{eq:precessioneq}) to first order in $w$ as
\begin{equation}
    \frac{d^2w}{d\phi^2} + w\underbrace{\left[1+3ku_c-s^{00}\frac{k^2}{2}\left(\frac{1}{2L_0^2}+3u_c^2\right)\right]}_{\equiv ~\Omega^2} \approx 0,
\end{equation}
which is an undamped harmonic oscillator with characteristic angular frequency $\Omega$. From this we identify the precession as $\Delta\phi = 2\pi(\Omega^{-1} -1)$, which reads
\begin{equation}
    \label{eq:deltaphi1}
    \begin{aligned}
        \Delta\phi =& 2\pi \left( 1+3ku_c-s^{00}\frac{k^2}{2}\left(\frac{1}{2L_0^2}+3u_c^2\right) \right)^{-\frac{1}{2}} -2 \pi\\
        \approx& -3 \pi ku_c + s^{00} \frac{\pi k^2}{2}\left(\frac{1}{2L_0^2}+3u_c^2\right) .
    \end{aligned}
\end{equation}

\subsection{Elliptic orbits far from the Schwarzschild radius}

We contrast the above result for nearly circular orbits with a similar result obtained through a different expansion, and thus valid in a different range of eccentricity and semi-major axis; we use an expansion for $u$ as $u = u_N + \delta u$, where $u_N=(1+e\cos{\phi})/p$ is the well-known elliptical solution to the harmonic oscillator, and where $e$ is the eccentricity and $p$ is the semi-latus rectum. From Eq.~(\ref{eq:precessioneq}) we find
\begin{equation}
    \begin{aligned}
        &\frac{d^2u_N}{d\phi^2}+u_N = -\frac{k}{2 L^2}, \\
        &L^2\frac{d^2\delta u}{d\phi^2} + \delta u\left(L^2-s^{00}\frac{k^2}{4}\right)-s^{00}\frac{k^2}{4}u_N -(u_N^2\\&+2u_N\delta u)\frac{3kL^2}{2} +(3u_N^2\delta u+u_N^3)s^{00}\frac{k^2L^2}{2} + \mathcal{O}(\delta u)^2 = 0.
    \end{aligned}
\end{equation}
We then separate $\delta u$ as $\delta u = \delta u_0 + \delta u_1$, where we consider $\delta u_1 / \delta u_0 = O(s^{00})$. We obtain two differential equations
\begin{equation}
    \label{eq:deltau1}
    \begin{aligned}
        &L^2\frac{d^2\delta u_0}{d\phi^2} + \left(L^2 - 2u_N \frac{3kL^2}{2} \right) \delta u_0 -u_N^2 \frac{3kL^2}{2} = 0, \\
        &L^2\frac{d^2\delta u_1}{d\phi^2} + \left(L^2 - 2u_N \frac{3kL^2}{2} \right) \delta u_1 - s^{00}\frac{k^2}{4} \delta u_0 \\& -s^{00}\frac{k^2}{4}u_N  + s^{00}\frac{k^2L^2}{2} (3u_N^2\delta u_0 +u_N^3)  = 0,
    \end{aligned}
\end{equation}
where $u_N \sim 1/p$, which we assume is small compared to the Schwarzschild radius $k/p \ll 1$ (terms such as $(k/p)^2$ or $k/p \delta u$ are therefore not considered); moreover, from the first equation of (\ref{eq:deltau1}), we know that $\delta u_0 \propto 1/p^2$. Given this, we can simplify (\ref{eq:deltau1}) as
\begin{equation}
    \begin{aligned}
        &\frac{d^2\delta u_0}{d\phi^2} + \delta u_0 = u_N^2 \frac{3k}{2}, \\
        &\frac{d^2\delta u_1}{d\phi^2} + \delta u_1 = s^{00}\frac{k^2}{4 L_0^2} u_N,
    \end{aligned}
\end{equation}
which we solve to obtain
\begin{equation}
    \begin{aligned}
        \delta u_0 =& - \frac{3 k}{2 p^2} \left( 1 + \frac{e^2}{2} + e \varphi \sin{\varphi} - \frac{e^2}{6} \cos{2 \varphi} \right) + c_1 \cos{\varphi} + c_2 \sin{\varphi}, \\
        \delta u_1 =& s^{00} \frac{k^2}{16 L_0^2 p} ( 4 \cos^2{\varphi} + 2 e \cos^3{\varphi} + 2 e \varphi \sin{\varphi} + 4 \sin^2{\varphi} \\&+ e \sin{\varphi} \sin{2\varphi} ) + c_3 \cos{\varphi} + c_4 \sin{\varphi},
    \end{aligned}
\end{equation}
where $c_1$, $c_2$, $c_3$ and $c_4$ are integration constants. In this expression, only terms which are not periodic will influence the precession, and we therefore have


\begin{equation}
    \label{eq:precess1}
    \begin{aligned}
        u_N + \delta u \approx \frac{1}{p} \left(1 + e \cos{\left(1 +\frac{3k}{2 p} - s^{00} \frac{k^2 }{8 L_0^2} \right) \varphi}  \right),
    \end{aligned}
\end{equation}

where we observe that this expression is no longer $2\pi$-periodic. For each period, the periastron will precess by an angle $\Delta \varphi$, which we compute as
\begin{equation}
    \begin{aligned}
        \Delta \varphi = 2 \pi\left(\frac{1}{1 + \epsilon} - 1\right) \approx -2 \pi \epsilon + \mathcal{O}(\epsilon^2).
    \end{aligned}
\end{equation}
From Eq.~(\ref{eq:precess1}) we readily identify $\epsilon = 3k/(2 p^2) - s^{00} k^2/(8 L_0^2 p)$ and find

\begin{equation}
\label{eq:precess}
    \begin{aligned}
        \Delta \varphi \approx -\frac{3 \pi k}{p} + s^{00} \frac{\pi k^2}{4 L_0^2} .
    \end{aligned}
\end{equation}

We find close agreement to  Eq.~(\ref{eq:deltaphi1}). We only miss a term that should resemble $3 \pi k^2 s^{00}/2u_c^2$, in order for the two expressions to  match in the limit of circular orbits $e \to 0$. We also note that the first term in $\Delta \varphi$ is a GR effect, and the other arises from symmetry breaking. We note that in addition to being suppressed by $s^{00}$, the second term is further inversely proportional to the square of angular momentum. We therefore posit that the most promising system for testing this effect is one in a strong gravity environment with small separations: recently, the GRAVITY collaboration released \cite{GRAVITY:2020gka,abuter2020detection} which shows Schwarzschild-like precession of the star S2 orbiting Sgr A* at the center of the Milky Way; such measurements can be used to place constraints on $s^{00}$ above, which we compute here for Case 1.

The ratio of the observed precession to that predicted by GR is, to $1\sigma$ significance
\begin{equation}
    \begin{aligned}
        f_{sp} = \frac{\Delta \varphi_{S2}}{\Delta \varphi_{GR}} = 1.10 \pm 0.19,
    \end{aligned}
\end{equation}
where we find that for Case 1, we have (using Eq.~(\ref{eq:precess}))
\begin{equation}
\label{eq:fsp}
    \begin{aligned}
        f_{sp} = 1 - s^{00} \frac{k p}{12 L_0^2}.
    \end{aligned}
\end{equation}

\begin{figure}
    \centering
    \begin{subfigure}[T]{0.49\textwidth}
        \includegraphics[width=\textwidth]{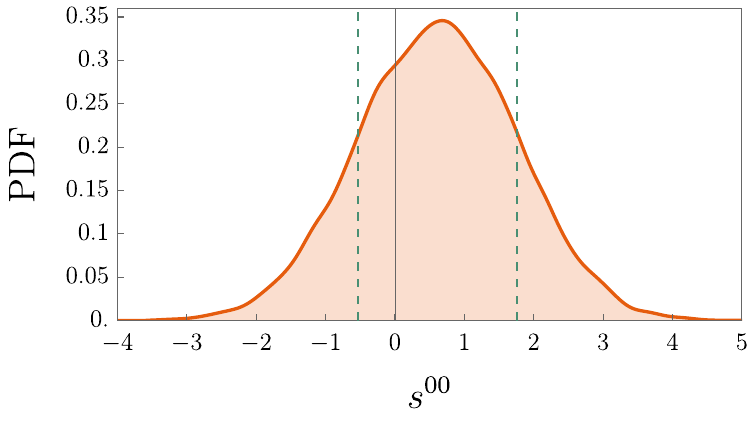}
        \caption{Posterior distribution function of $s^{00}$ estimated from the S2 motion as reported by \cite{abuter2020detection}, where dashed green lines represent $1\sigma$ confidence level.} 
        \label{fig:s00posterior}
    \end{subfigure}
    \hfill
        \begin{subfigure}[T]{0.49\textwidth}
        \includegraphics[width=\textwidth]{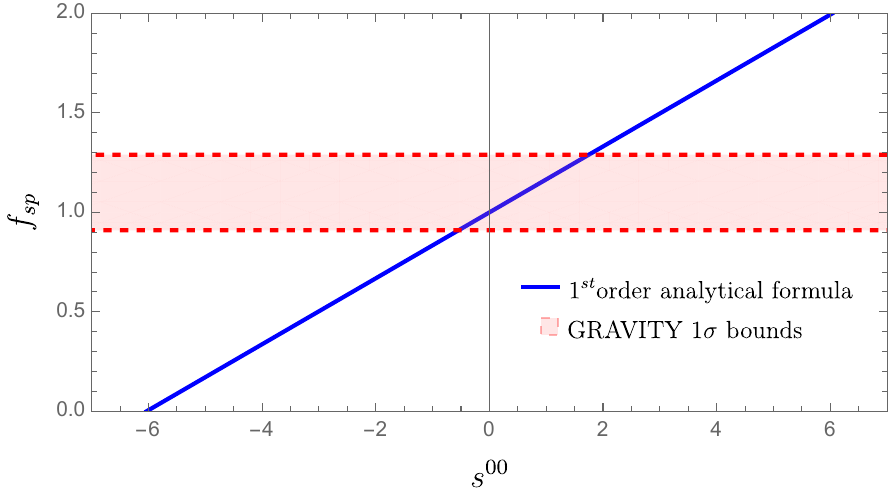}
        \caption{Represented is $f_{sp}$ with respect to $s^{00}$ through (\ref{eq:fsp}). The physical parameters $k$, $p$ and $L^2$ are those of the orbit of S2 around Sgr A*.} 
        \label{fig:fspbounds}
    \end{subfigure}
    \caption{Estimated posterior distribution function for $s^{00}$ (left); comparison between $f_{sp}$ and analytical solution for $s^{00}$ using best-fit parameters (right).}
\end{figure}

For a given bound trajectory of a massive object, we may compute the constants $E^2$ and $L^2$ through Eq.~(\ref{eq:energ1}), as well as the identities $dr/d\tau|_{r_a} = dr/d\tau|_{r_p} = 0$, where $r_a$ and $r_p$ are the radius of the apoastron and periastron, respectively. We find that
\begin{equation}
    \begin{aligned}
        L^2 &= r_p^2 r_a^2 \frac{N_a^2 - N_p^2}{r_a^2 N_p^2 - r_p^2 N_a^2}, \\
        E^2 &= N_p^2 N_a^2 \frac{r_a^2 - r_p^2}{r_a^2 N_p^2 - r_p^2 N_a^2},
    \end{aligned}
\end{equation}
where $N_a$ and $N_p$ represent the metric function $N(r)$ evaluated at $r=r_a$ and $r=r_p$, respectively. For the values of the system, we adopt the $1\sigma$ bounds obtained in \cite{GRAVITY:2020gka} as $a=125.4\pm 0.018$ mas, $e=0.88466\pm0.00018$, and $M_{SgrA*}=4.100\pm0.034 \times 10^6 \, M_{\odot}$; moreover, we take the distance between SgrA* and the Earth we take $R_0=8122\pm31$ pc. In order to estimate the resulting distribution for $s^{00}$, we assume Gaussian $1\sigma$ bounds on the above parameters and generate posteriors with $10^3$ mock data points, after which we can estimate $s^{00}$ through Eq.~(\ref{eq:fsp}) as
\begin{equation}
\label{eq:fspboundsval}
    \begin{aligned}
       \boxed{s^{00} = 0.62^{+1.14}_{-1.16} \,\, (1\sigma)},
    \end{aligned}
\end{equation}
the posterior of which can be seen in Figure~\ref{fig:s00posterior}.
Another option is to simply adopt the central values for the system parameters \{$a$, $e$, $M_{SgrA*}$, $R_0$\} and use the $1\sigma$ bounds on $f_{sp}$ above to find $s^{00}=0.60\pm 1.15$. We also show the dependence of $s^{00}$ in Eq.~(\ref{eq:fsp}) (using the best-fit system parameters) together with the $1\sigma$ bound on $f_{sp}$ in Figure~\ref{fig:fspbounds}.

Compared to constraints obtained elsewhere in the gravity sector, these bounds are quite weak; for example, a constraint on $s_{00}$ was found from the speed of gravity measurement using the gravitational-wave event GW170817 with associated kilonova counterpart GRB170817A as $-6\cdot 10^{-15}<s_{00}<+1.4\cdot 10^{-14}$. It is worth noting that due to the explicit-breaking nature considered here $s^{\mu\nu}$ and $s_{\mu\nu}$ should be considered different theories. Here, we have computed a match between the GRAVITY results and translated the bounds to $s^{00}$, assuming Gaussian posteriors; a computation of a more precise constraint is currently underway.

\section{Light ring}\label{sec:lightring}
We can define light rings, a special class of null geodesics, for our Schwarzschild-like solutions. Generally, light rings are spanned by null vectors tangent to a combination of $K_t^\mu$ and $K_\phi^\mu$ (see Section~\ref{sec:consquantsorbits}), and can be classified as being either stable or unstable under perturbations~\cite{Cunha:2018acu}. Unstable light rings (for example, all light rings for a Kerr black hole) are important in the formation of the black-hole shadow, where a photon may circle the black hole for some time before falling inside the horizon or scattering to infinity. Stable light rings can appear around for example the Schwarzschild black holes and Proca stars~\cite{Cunha:2017wao}. 

\subsection*{Radius}
Given the similarity of our solutions to Schwarzschild and Reissner-Nordstr\"om, we pick Case 1 and derive the effect on the light ring radius; we leave the more exotic case for future work.

To calculate this radius, one can compare the geodesic equations for $d^2 r/d \tau^2$, and the line element $ds^2$ under the assumptions of a circular orbit.
\begin{equation}
    \begin{aligned}
        \frac{d^2 r}{d \tau^2} = \frac{d r}{d \tau} = 0.
    \end{aligned}
\end{equation}
Since we have spherical symmetry, we may choose $\theta = \pi/2$ without loss of generality. The geodesic equation for $r$ appears as
\begin{equation}
    \begin{aligned}
        \frac{d^2 r}{d \tau^2} &= -\Gamma^{1}_{\mu\nu} u^\mu u^\nu = - \Gamma^{1}_{00} \left( \frac{d t}{d \tau} \right)^2 - \Gamma^{1}_{33} \left( \frac{d \phi}{d \tau} \right)^2 = 0.
    \end{aligned}
\end{equation}

Introducing the values $\Gamma^{1}_{00}$ and $\Gamma^{1}_{11}$ with respect to $N$ and $f$ and their expressions in (\ref{eq:scase1}), we find
\begin{equation}
\label{eq:lr1}
    \begin{aligned}
        \left( \frac{d \phi}{d t} \right)^2 = - \frac{k}{2 r_{lr}^3} + s^{00} \frac{k^2}{4 r_{lr}^4} + s^{00} \frac{5 k^3}{8 r_{lr}^5},
    \end{aligned}
\end{equation}
where $r_{lr}$ is the radius of the light-ring. Through the line element of a photon $ds^2 = 0$, we get another expression for the derivative of $\phi$ with respect to $t$
\begin{equation}
\label{eq:lr2}
    \begin{aligned}
        \left( \frac{d \phi}{d t} \right)^2 = \frac{1}{r_{lr}^2} + \frac{k}{r_{lr}^3} - s^{00} \frac{k^2}{4 r_{lr}^4}.
    \end{aligned}
\end{equation}

Equaling the two expressions (\ref{eq:lr1}) and (\ref{eq:lr2}), we have
\begin{equation}
\label{eq:lr3}
    \begin{aligned}
        1 + \frac{3k}{2 r_{lr}} - s^{00} \frac{k^2}{2 r_{lr}^2} - s^{00} \frac{5k^3}{8 r_{lr}^3} =0,
    \end{aligned}
\end{equation}
and thanks to our perturbative formulation of $r_{lr}$ around the Schwarzschild solution and the expansion (\ref{eq:dvlim}), we find
\begin{equation}
\label{eq:lr4}
\boxed{
    \begin{aligned}
        r_{lr} = -\frac{3}{2} k \left(1 + \frac{s^{00}}{27} \right) + \mathcal{O}((s^{00})^2),
    \end{aligned}}
\end{equation}
where the Schwarzschild light ring at $r=3M$ is recovered when $s^{00} \to 0$.

\subsection*{Stability}
Given that the Schwarzschild light ring is unstable, we can expect that the same holds for Eq.~(\ref{eq:lr4}); we now check this by explicit computation. 

From (\ref{eq:energycons}) with $\epsilon = 0$ we find
\begin{equation}
\label{eq:energyconsphot}
\begin{aligned}
    &\frac{d^2r}{d\tau^2} = 3\frac{kL^2}{2r^4} - s^{00}\frac{k^2L^2}{2r^5} + \frac{L^2}{r^3},
\end{aligned}
\end{equation}
and we note that determining the sign of $d^2 r/d \tau^2$ is equivalent to finding the sign of  the polynomial $2 \Bar{r}^2 + 3k \Bar{r} - s^{00} k^2$. The discriminant $\Delta = k^2 \left(9 + 8 s^{00} \right)$ is always strictly positive since $s^{00} \ll 1$. The roots $\Bar{r}_{1,2}$ of this polynomial are
\begin{equation}
\label{eq:energyconsphot2}
\begin{aligned}
    &\Bar{r}_{1} = \frac{k}{3} s^{00} + \mathcal{O}((s^{00})^2) \\
    &\Bar{r}_{2} = -\frac{3k}{2} \left(1 + \frac{2}{9} s^{00} \right) + \mathcal{O}((s^{00})^2).
\end{aligned}
\end{equation}
From the sign of the coefficient in front of $\Bar{r}^2$, (\ref{eq:lr4}) corresponds to a maximum of $dr/d \tau$ and the light-ring is unstable.

We leave the question of the black-hole shadow for future work.

\section{Discussion \& Conclusions}\label{sec:disc}
In this paper we found and investigated four spherically-symmetric vacuum solutions in the presence of a non-dynamical background field in the form of a two-tensor. More precisely, we studied a subclass of a general EFT construction where particle diffeomorphism symmetry is explicitly broken. By allowing different components of the diagonal elements to be non-zero, we constructed perturbative solutions in a spherically-symmetric and static geometry, thus obtaining the analytical expressions of different phenomena that can readily be confronted to observations. Indeed, the observational status of static and spherically symmetric spacetimes is evolving : in \cite{Bonder:2020fpn}, the authors consider explicit breaking using the second term in our Eq.~(\ref{eq:riccidecomp}), using which they obtain perturbative solutions. Using bounds from the GREAT experiment they obtain constraints on the components of the EFT coefficients. Although based on the same EFT, our solutions are not contained in these results as they assume traceless coefficients.
More recently, bumblebee black holes were tested using data from the Event Horizon Telescope, where it was found that the parameter space is largely unexplored \cite{Xu:2023xqh}. Although the bumblebee model is a vector subset of the action (\ref{eq:lagr}), symmetry breaking happens spontaneously and is therefore not a subset of the results considered in the present manuscript. \\
We investigated here the properties of our solutions' horizons, thermodynamics, and singularities, where we found that several of the models have a positive peak followed by a negative divergence in the Kretschmann scalar as $r\to 0$. We derived the behavior of radial photon geodesics and determined the presence of stable/unstable circular orbits reminiscent of the Schwarzschild solution. Finally, we derived the precession of the periastron for one of our new solutions. The quadratic corrections at linear order in the EFT coefficients make these solutions similar to the Reissner-Nordstr{\"o}m case, with the Reissner-Nordstr{\"o}m charge proportional to $s^{00} k^2$ for Case 1 and 2, both of which contain only one non-zero component of the coefficient $s^{\mu\nu}$. Similarly to Reissner-Nordstr{\"o}m, these solutions show an outer event horizon as well as an internal Cauchy horizon. Strikingly, we find that when introducing a second non-zero coefficient in Case 3, we obtain a solution with {\it three} horizons. It is interesting to consider whether the appearance of additional horizons is an artefact of the perturbative nature of the solutions. Indeed, similar spurious effects, which do not have a well-defined LI limit, have been known to appear in the context of modified dispersion relations in the presence of Lorentz violation. This can also be seen in the asymptotic series used for post-Minkowskian/Newtonian calculations, as well as in EFT approaches explicit diffeomorphism breaking, where discontinuities have been found~\cite{Bailey:2024zgr}. In the present context of black-hole solutions, additional horizons would appear also in an exact solution, as can be demonstrated simply: the EFT term in the action~(\ref{eq:lagr}) can be schematically written as $s \cdot R \sim s g^\prime g^\prime$. Taking $g$ to be of Schwarzschild form $k/r$, we have that $s g^\prime g^\prime \sim s/r^4$ after differentiation. It is therefore not surprising to obtain solutions with more than two horizons, for example those of RN-type ($\sim 1/r^2$) or even more complicated ones where $s^{11} \neq0$ in addition to $s^{00}$.

For Case 1 and the trace model, we find that the Hawking temperature evolves similarly to the Schwarzschild case as a function of the horizon radius for $s^{00}<0$ with a divergence as $r_\star \to 0$; however, for $s^{00}>0$ we find a peak in $T_H$ followed by a negative divergence in the temperature. The behavior for Cases 2 and 3 is richer, with both positive and negative singularities appearing at the Schwarzschild horizon as well as for small $r_\star$. More interesting behavior is found in the Kretchmann scalar, where we find that the singularity at $r=0$ is preceded by a maximum at small $r$, outside the inner horizon (for Case 1 and the trace case), whereas Cases 2 and 3 show a curvature singularity at the Schwarzschild horizon. Moreover, the Ricci scalar is distinctly non-zero for Case 2 and 3, a significant difference from GR; in this sense, Case 1 and the trace case are less exotic solutions. Given the $1/r^2$ corrections arising from the EFT coefficients, these solutions all resemble Reissner-Nordstr\"{o}m type metrics. The exception is Case 3: given that we find three horizons, we expect this solution to have behavior different from that of the others cases considered here; however, the Hawking temperature and singularity structure is, while not identical, not significantly different from Case 2. Black hole solutions with more than two horizons are known in the literature, where they, for example, appear as Reissner-Nordstr\"{o}m de Sitter solutions~\cite{Romans:1991nq}, as well in the context of Loop Quantum Gravity~\cite{Kumar:2022vfg} and gravity with non-linear electrodynamics~\cite{Nojiri:2017kex}. Although such solutions have the same number of horizons as our Case 3, they bear little resemblance, as the outer horizon generally sits at infinity (the cosmological horizon, which is also present in the Schwarzschild de Sitter solution). 
We also note that for both Case 2 and 3, curvature singularities and Hawking temperature divergencies persist at the Schwarzschild horizon, which in both cases is {\it not} a horizon for these solutions; instead, the inner and outer horizons sit symmetrically on either side of the Schwarzschild counterpart. This is in contrast to several solutions in the literature. Since all horizons coincide for $s^{\mu\nu}\to0$, the Schwarzschild solution can be seen as the extremal limit of these more general solutions. 

Another point worth noting is the appearance of naked singularities in Case 2. Given the weak Cosmic Censorship Conjecture~\cite{1969NCimR...1..252P}, no singularities should be visible to an observer at future null infinity ($\mathcal{I}^+$), and as such, this solution ought to be forbidden, as the spacetime manifold is no longer strongly asymptotically predictable. However, there are plenty of examples of naked singularities in the literature, even in General Relativity, such as the extremal Reissner-Nordstr\"om solution. In scalar-tensor theories, black-hole solutions with primary scalar hair have been shown to possess naked singularities~\cite{Bakopoulos:2023fmv}, and one can create classes of solutions with mild assumptions on the metric functions $N(r)$ and $f(r)$. One may expect that such solutions are unstable, which requires a perturbation treatment; we leave this for future work.

In regards to photon radial geodesics, we found that corrections in powers of $(r+k)^{-1}$ appear when considering Cases 1, 2, and 3. For Case 1, we retain the standard property of radial photon geodesics, i.e. $ct/r \to 1$ when $r \to \infty$; however, this is not the case for Case 2 and 3, where the presence of $s^{11}$ in the term linear in $r$ in (\ref{eq:radgeos11}) and (\ref{eq:radgeoall}) implies that $ct/r \to 1 + s^{11}/2$ at infinity.
The study of the behaviour of mechanical energy in (\ref{eq:energycons}) and (\ref{energyphotcase2}) showed that depending on the sign of $s^{00}$ and presence of $s^{11}$, a potential barrier may appear at a strictly positive radius, thus forbidding massive particles from penetrating past a certain hypersurface.
Additionally, we calculated the periastron precession for quasi-circular orbits, as well as more eccentric orbits. Using these results, we placed some rudimentary constraints on the EFT coefficient in Case 1 through the value of the parameter $f_{sp}$ describing the ratio between the observed precession and the predicted precession of GR, with S2 orbit (see~\cite{GRAVITY:2020gka}): $-0.54 < s^{00} < 1.76 $. These constraints may not yet be very competitive, but they illustrate the potential of this work when confronted with observational data.
Finally, we studied the presence of a light-ring around the black hole and found that one exists, albeit with a slightly perturbed radius with respect to Schwarzschild, and that it remains unstable (as in Schwarzschild).

In conclusion, the solutions presented in this paper are meant to act as a tool to study diffeomorphism invariance violation in the static and spherically symmetric spacetime that we can observe around a non-rotating black hole. We will use them as such in future work, in which they will be confronted to observational data from a number of missions both ground and space-based, for example EHT and GRAVITY.

\begin{acknowledgements}
The authors thank Quentin G. Bailey and Theodoros Nakas for useful comments. N.A.N. was financed by CNES and IBS under the project code IBS-R018-D3, and acknowledges support from PSL/Observatoire de Paris.
\end{acknowledgements}

\appendix

\vspace{5mm}

\begin{center}
    \textbf{Appendix}
    \vspace{-5mm}
\end{center}

\section{Killing horizons}\label{app:killinghorizons}
The Killing horizons of a static and spherically symmetric solution (of radius $r_{\star}$) must respect two conditions:
({\it i}) the hypersurface must be null; and
({\it ii}) the Killing vector $K_t = \partial_t$ (associated with the stationary isometry) must be orthogonal to it. However, we have calculated $f_1$ and $N_1$ in Schwarzschild coordinates $(t, r, \theta, \varphi)$ in which the metric is diagonal, and this results in any Killing horizon $g_{tt} (r_{\star}) = 0$ also being a coordinate singularity $g^{tt} (r_{\star}) = \pm \infty$. In order to resolve this ambiguity, we show below that there exists a coordinate system $(\Bar{t}, r, \theta, \phi)$ on the horizon where we have that
\begin{equation}
\label{eq:cdtcoord}
    \begin{split}
        &K_t = \partial_{\bar{t}}, \\
        &K_t \cdot K_t = g_{\bar{t} \bar{t}} = 0, \\
        &K_t \cdot \partial_\theta = g_{\bar{t} \theta} = 0,\\
        &K_t \cdot \partial_\phi = g_{\bar{t} \phi} = 0,\\
        &g^{\alpha\beta} \neq \pm \infty,
    \end{split}
\end{equation}
which are necessary and sufficient conditions for the existence of a Killing horizon. The new coordinate system $(\Bar{t}, r, \theta, \varphi)$ can be seen as analogous to the Eddington-Finkelstein coordinates.

\subsection{Case 1}
Using the transformation
\begin{equation}
    \begin{split}
        &t \to \Bar{t} = t + h_{1}(r),\\
        &h_{1}(r) = -k\ln{\left| \frac{r}{k} + 1\right|} + k\frac{s^{00}}{4} \left(\frac{r}{k} + 1 \right)^{-1},
    \end{split}
\end{equation}
we find the metric as
\begin{equation}
\label{eq:case1edd}
    \begin{split}
        \mathrm{d}s^2 = &-\left( 1 + \frac{k}{r} - s^{00} \frac{k^2}{r^2} \right) \mathrm{d}\bar{t}^2 - \left( \frac{2k}{r} - s^{00} \frac{k^2}{2 r^2}  \right) \mathrm{d}\bar{t} \mathrm{d}r \\
        &+ \left(1 - \frac{k}{r} + s^{00} \frac{k^2}{4r^2} \right)\mathrm{d}r^2 + r^2 \left( \mathrm{d}\theta^2 + \sin^2{\theta} \mathrm{d}\varphi^2 \right).
    \end{split}
\end{equation}
This new metric is regular on the roots of $N^2(r)=0$ in its covariant and contravariant components, and reduces to Eddington-Finkelstein coordinates when $s^{00} \to 0$. It only diverges when $r=0$, but thanks to the Kretschmann scalar (\ref{eq:case1scalars}), we know that this corresponds to a "true" curvature singularity. Since the Killing horizon corresponds to the roots $r_{KH}$ of $N^2 (r_{KH}) = 0$, we have for Case 1 that $r_{KH} = r_{\star}$.

\subsubsection{Trace Case}
Let us consider the following coordinate transformation for the trace term
\begin{equation}
\label{eq:casetraceedd}
    \begin{split}
        &t \to \bar{t} = t + h_{1, trace}(r),\\
        &h_{1, trace}(r) = -k \left(1 + s^{00} \right) \ln{\left| \frac{r}{k} + 1 \right|} \\&+ k\frac{s^{00}}{2} \left( \frac{r}{k} + 1 \right)^{-1}.
    \end{split}
\end{equation}
From this, we obtain the metric tensor in the new coordinates as
\begin{equation}
    \begin{split}
        \mathrm{d}s^2 = &-\left( 1 + \frac{k}{r} + s^{00} \left( \frac{3k^2}{2r^2} - \frac{2k}{r} \right) \right) \mathrm{d}\bar{t}^2 \\
        &- \left( \frac{2k}{r} - s^{00} \left( \frac{3k^2}{r^2} + \frac{2k}{r} \right)  \right) \mathrm{d}\bar{t} \mathrm{d}r \\
        &+ \left(1 - \frac{k}{r} - s^{00} \frac{3k^2}{2r^2} \right)\mathrm{d}r^2 \\&+ r^2 \left( \mathrm{d}\theta^2 + \sin{\left(\theta \right)}^2 \mathrm{d}\varphi^2 \right),
    \end{split}
\end{equation}
which only diverges on the curvature singularity $r=0$, same as for Case 1 (see Eq.~\ref{eq:case1edd}); however, the identity $N^2 = f + \mathcal{O}\left(s^{11})^2 \right)$ does not hold in the Trace Case 1 and as a consequence $r_{KH} \neq r_{\star}$. The radius of the Killing horizon therefore has to be calculated from the identity $N^2(r_{KH}) = 0$; doing so, one obtains
\begin{equation}
\label{eq:khtrace}
    \begin{split}
        r_{KH,trace} =& - \frac{k}{2} \left( 1 + s^{00} \pm \left( 1 - \frac{s^{00}}{2} \right) \right) \\&+ \mathcal{O} \left((s^{00})^2 \right).
    \end{split}
\end{equation}
They are quite similar to the horizons of Case 1 in Eq.~(\ref{eq:case1hor}), as one could expect from the divergence of the metric (see Eqs~\ref{eq:scase1} and \ref{eq:tracesol}). These are the only radii at which one could have a Killing horizon linked to $\bm{K}_t$ in the trace case.

\subsection{Case 2}
Let us consider the coordinate change
\begin{equation}
    \begin{split}
        t \to \bar{t} =& \: t + h_{2}(r),\\
        h_{2}(r) =& - k \left(1 - s^{11} \frac{11}{16} \right) \ln{\left| \frac{r}{2M} -1 \right|} - s^{11} \frac{k}{2} \left( \frac{r}{k} + 1 \right)^{-1} \\
        &+ s^{11} \frac{k}{16} \left( \frac{r}{k} + 1 \right)^{-2}.
    \end{split}
\end{equation}
In this new coordinate system the line element $ds^2$ reads
\begin{equation}
\label{eq:case2edd}
    \begin{split}
        \mathrm{d}s^2 = &- \left( 1 + \frac{k}{r} + s^{11} \frac{k}{8r} \left( 4 + 5 \frac{k}{r} \right) \left( 1 + \frac{k}{r} \right)^{-1} \right) \mathrm{d}\bar{t}^2 \\
        &- \frac{2k}{r} \left(1 - s^{11} \frac{11}{8} \right) \mathrm{d}\bar{t} \mathrm{d}r \\
        &+ \biggl( 1 - \frac{k}{r} - s^{11} \biggl( 1 - \frac{2k}{r} + \frac{7}{8} \left( \frac{r}{k} + 1 \right)^{-2} \\
        &+ \frac{7}{8} \left( \frac{r}{k} + 1 \right)^{-3} \biggr) \biggr) \mathrm{d}r^2 + r^2 \left( \mathrm{d}\theta^2 + \sin{\left(\theta \right)}^2 \mathrm{d}\varphi^2 \right),
    \end{split}
\end{equation}
which indicates that it is not possible to find a function $h_2 (r)$ such that the metric is only singular at $r = 0$ but not in $r=-k$, such that the new coordinates reduce to Eddington-Finkelstein coordinates when $s^{11}\to 0$. This feature of the metric seems to confirm what the analysis of the curvature invariants (see Figure~\ref{fig:case3curvature}) showed, namely the presence of a true curvature singularity in $r=-k$. Nevertheless, we constructed these coordinates such that the metric would {\it not} be singular on $r_{KH}$, and that is accomplished. Thanks to this coordinate system we can see without ambiguity that the roots of $N^2 (r_{KH}) = 0$ is the radius of a Killing horizon, and not a coordinate singularity.

Since $N^2 \neq f$, we calculate the roots of $N^2 (r_{KH}) = 0$ and find
\begin{equation}
\label{eq:khcase2}
    \begin{split}
        &r_{KH,2} = -k \left(1 + \frac{s^{11}}{2} \pm \sqrt{-\frac{1}{4} s^{11}} \right) + \mathcal{O}(( s^{11})^2),
    \end{split}
\end{equation}
which as for the trace case (\ref{eq:khtrace}) differ only slightly from the solutions outlined in (Eq.~\ref{eq:case2hor}).

\subsection{Case 3}
Since our construct is linear in the coefficients $s^{\mu\nu}$, the appropriate coordinate transformation for Case 3 is the sum of Case 1 and Case 2, which reads
\begin{equation}
    \begin{split}
        t \to \bar{t} =& \: t + h_{3}(r), \\
        h_{3} (r) =& \: -k \left( 1 + s^{00} - \frac{11}{16} s^{11} \right) \ln{\left| \frac{r}{2M} -1 \right|} \\
        &+ \frac{k}{2} \left( s^{00} - s^{11}  \right) \left( \frac{r}{k} + 1 \right)^{-1} + \frac{k}{16} s^{11} \left( \frac{r}{k} + 1 \right)^{-2}.
    \end{split}
\end{equation}
This leads to a new line element that is the combination of Eqs~(\ref{eq:case1edd}) and (\ref{eq:case2edd})
\begin{equation}
\label{eq:case3edd}
    \begin{split}
        \mathrm{d}s^2 = &- \left( 1 + \frac{k}{r} - s^{00} \frac{k^2}{r^2} + s^{11} \frac{k}{8r} \left( 4 + 5 \frac{k}{r} \right) \left( 1 + \frac{k}{r} \right)^{-1} \right) \mathrm{d}\bar{t}^2 \\
        &- \left( \frac{2k}{r} \left(1 - \frac{11}{8} s^{11} \right) - s^{00} \frac{k^2}{r^2} \right) \mathrm{d}\bar{t} \mathrm{d}r \\
        &+ \biggl( 1 - \frac{k}{r} + s^{00} \frac{k^2}{4 r^2} - s^{11} \biggl( 1 - \frac{2k}{r} + \frac{7}{8} \left( \frac{r}{k} + 1 \right)^{-2} \\
        &+ \frac{7}{8} \left( \frac{r}{k} + 1 \right)^{-3} \biggr) \biggr) \mathrm{d}r^2 + r^2 \left( \mathrm{d}\theta^2 + \sin{\left(\theta \right)}^2 \mathrm{d}\varphi^2 \right).
    \end{split}
\end{equation}
As for the other Cases, we solve $N^2 (r_{KH}) = 0$ and obtain the radii of the Killing horizons as
\begin{equation}
\label{eq:khcase3}
    \begin{cases}
        r_{KH,3} &= s^{00} \frac{k}{4}, \\
        r_{KH,3} &= -k \left( 1 + \frac{s^{11}}{2} + \frac{s^{00}}{8} \pm \sqrt{-\frac{s^{11}}{4}} \right),
    \end{cases}
\end{equation}
which is similar to the direct sum of the solutions (\ref{eq:case1hor}) and (\ref{eq:khcase2}).

\section{Simple exact solutions}\label{app:simpleexact}
\subsection{Case 1}
\label{subsec:appcase1sol}
Looking for trivial solutions of Eq.~(\ref{eq:ELcase1}) which are non-linear differential equations, we take $N=\text{const.}$, plug it in (\ref{eq:ELcase1}) and find that we necessarily must have $f = N = 1$. If we instead take $f=\text{const.}$, there are still two non-linear differential equations for $N$ that cannot be trivially reduced to one, making it hard to study their solution space.

\subsection{Case 1 Trace}
\label{subsec:appcase1tracesol}
When $N$ is taken to be constant in (\ref{eq:tracesol}), we obtain the solution 
\begin{equation}
    f(r) = 1 + \frac{\alpha}{r},
\end{equation}
where $\alpha$ is an integration constant. If $\alpha=0$ then $N$ is completely undetermined, but for $\alpha\neq0$ then the solution $N = -1/s^{00}$ is forced by the equation. As in Case 1, when $f$ is taken to be constant, we obtain two non-linear differential equation for $N$ that are non-trivially related.

\subsection{Case 2}
\label{subsec:appcase2sol}
The Euler-Lagrange equations for $N$ and $f$ read
\begin{equation}
    \label{eq:case2eullag}
    \begin{split}
    &0 = 4 f^3 + 3r^2 s^{11} f'^2 + 4f^2 \left( -1 + s^{11} + r f' \right) \\&- 2 r s^{11} f \left( 2f' + r f'' \right) \\
    &0 = N \left(-1 + f + s^{11} \right) + r \big( 2 f N' + 2 s^{11} N' \\&+ r s^{11} N'' \big)
    \end{split}
\end{equation}
If we impose the condition $f'=f''=0$, we find immediately that $f= 1- s^{11}$. Plugging this result into the second equation above, we find a linear differential equation for $N$ which solves as
\begin{equation}
    N(r) = \alpha + \beta r^{\gamma},
\end{equation}
where $\alpha$ and $\beta$ are integration constants, and $\gamma = 1 - 2(s^{11}-2)/s^{11}$. When $|s^{11}|\ll1$, $\gamma$ is of the order of $4/s^{11}$.
If $N$ is a constant and $N\neq 0$, then $f$ is also a constant and $f=1-s^{11}$. If $N=0$, then $f$ is still defined by the first non-linear differential equation in (\ref{eq:case2eullag}).

\vspace{5mm}


\bibliography{apssamp}

\end{document}